\documentclass{article}
\usepackage{arxiv}
\usepackage{graphicx}
\usepackage{amsmath}
\usepackage{tabularx}
\PassOptionsToPackage{hyphens}{url}\usepackage{hyperref}
\usepackage[all]{hypcap}
\usepackage[misc]{ifsym}
\usepackage{color, soul}
\usepackage{calc}
\usepackage{empheq}
\usepackage[many]{tcolorbox}
\tcbset{
  myhlight/.style={
    colback=yellow, 
    arc=0pt,
    outer arc=0pt,
    boxrule=0pt,
    top=2pt,
    bottom=2pt,
    left=2pt,
    right=2pt,
  },
  highlight math style={myhlight}
}

\newtcolorbox{myhl}{
  breakable,
  myhlight
}
\begin{document}

\title{Resonating Minds -- Emergent Collaboration Through Hierarchical Active Inference} 




\author{\href{https://orcid.org/0000-0002-9677-2755}{\includegraphics[scale=0.06]{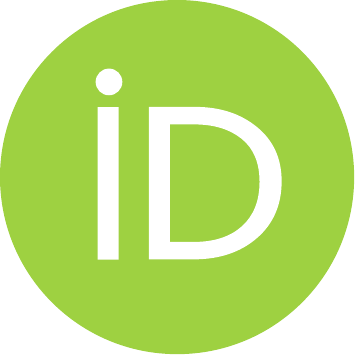}\hspace{1mm}Jan P\"oppel}\\
	Social Cognitive Systems\\
	Bielefeld University \\
	\texttt{jpoeppel@techfak.uni-bielefeld.de} \\
	\And
	 \href{https://orcid.org/0000-0002-8468-2808}{\includegraphics[scale=0.06]{orcid.pdf}\hspace{1mm}Sebastian Kahl}\\
	Social Cognitive Systems\\
	Bielefeld University \\
	\texttt{skahl@techfak.uni-bielefeld.de} \\
	\And
	\href{https://orcid.org/0000-0002-4047-9277}{\includegraphics[scale=0.06]{orcid.pdf}\hspace{1mm}Stefan Kopp}\\
	Social Cognitive Systems\\
	Bielefeld University \\
	\texttt{skopp@techfak.uni-bielefeld.de} \\
}

\maketitle

\begin{abstract}

\textbf{Introduction}
Working together on complex collaborative tasks requires agents to coordinate their actions.
Doing this explicitly or completely prior to the actual interaction is not always possible nor sufficient.
Agents also need to continuously understand the current actions of others and quickly adapt their own behavior appropriately.
Here we investigate how efficient, automatic coordination processes at the level of mental states (intentions, goals), which we call belief resonance, can lead to collaborative situated problem-solving. \\
\textbf{Method}
We present a model of hierarchical active inference for collaborative agents (HAICA). It combines efficient Bayesian Theory of Mind processes with a perception-action system based on predictive processing and active inference.
Belief resonance is realized by letting the inferred mental states of one agent influence another agent's predictive beliefs about its own goals and intentions.
This way, the inferred mental states influence the agent's own task behavior without explicit collaborative reasoning.\\ 
\textbf{Results}
We implement and evaluate this model in the Overcooked domain, in which two agents with varying degrees of belief resonance team up to fulfill meal orders.
Our results demonstrate that agents based on HAICA achieve a team performance comparable to recent state of the art approaches, while incurring much lower computational costs.
We also show that belief resonance is especially beneficial in settings were the agents have asymmetric knowledge about the environment. \\ 
\textbf{Conclusion}
The results indicate that belief resonance and active inference allow for quick and efficient agent coordination, and thus can serve as a building block for collaborative cognitive agents.

\end{abstract}


\section{Introduction}
Collaboration is a fundamental human skill. Many situations we are confronted with can only be mastered when working together with others on a common goal and in a coordinated effort. For example, when preparing a meal together in a kitchen, agents need to coordinate to determine sub-tasks for each agent or for jointly working on a sub-task when necessary. Often this requires explicit coordination prior to task execution, e.g., through jointly planning and negotiating sub-tasks. However, such a prior coordination is often not sufficient nor even possible, because most tasks take place in unpredictable environments or with agents that may behave in unexpected ways. Likewise, agents may not always be able to coordinate explicitly, e.g., by communicating, during the interaction.  
Consequently, agents need to coordinate their actions on-the-fly in order to collaborate successfully.

We humans have evolved rich abilities for coordinating our collaborative actions on-the-fly. In social interaction, we manage to work together even without prior experience or fixed social roles \cite{tomaselloUnderstandingSharingIntentions2005,misyakUnwrittenRulesVirtual2014}. This is possible because, despite a lack of direct access to other agents' minds, 
from a young age we are able to perceive the actions of other agents and form hypotheses about their current intentional state (action intentions, plans, or goals) or epistemic-attentional state (beliefs, assumptions, perceptions) \cite{gergely2003teleological}.  
The underlying ability is called mentalizing or \textit{Theory of Mind} (ToM) \cite{premack:1978does}. 
At a perceptual level, we are able to recognize and predict another agent's actions and motor intentions \cite{vesper:2017joint} by way of 
\emph{resonance} processes that involve our own motor system in our perception of others \cite{schutz:2007perceptual}.

Obviously, successful coordination of collaboration does not only require recognizing what other agents are currently doing, which intentions they might act upon, and what this may imply for the entire joint task. 
Collaborative agents also need to adapt their own actions accordingly in order to ensure coordinated behavior. 
While this may require explicit re-planning and even re-negotiating sub-tasks in some cases (e.g. \textit{``So you are getting the plate? Then I will get the cutlery.''}), it may also happen implicitly and continuously, e.g., in a situated joint action such as carrying a table together. How these two processes (explicit and implicit coordination) interact and play out together is a largely open question. The former is able to handle complex collaboration problems with rich representations of task and partner knowledge, but the cognitive inference and decision processes are known to be uncertain, costly and often even intractable \cite{blokpoel2013computational}; the latter is efficient and robust in dynamic environments and assumed to rest on sensorimotor processes, but it seems restricted to coordinative patterns of single motor actions. Finally, agents may even happen to collaborate without any deliberate coordination, e.g., when the environment restricts individual actions and in a way that brings about their coordination. 
 
In this paper we ask if, and to what extent, successful collaboration between artificial agents on complex situated tasks can emerge even from minimal, distributed on-the-fly coordination processes. Traditionally, collaboration in artificial agents has been approached as planning problem that must be solved either in a centralized or decentralized fashion \cite{de2009introduction}. 
Enabling complex collaborative tasks with real-time on-the-fly coordination between autonomous agents, however, is still a challenge and usually requires approximations or heuristic processes. 
One reason for this is that both generating accurate action predictions for other agents using a full-blown ToM and optimal planning in complex, dynamic environments are too costly to be performed online \cite{baker2009action,albrecht2013ad}.
Simplifications are thus commonplace. It is for this reason that, e.g., most artificial agents perform best in simulations in which the other agents are very similar to themselves (as found in \cite{carroll2019utility}).

We propose an inter-agent coordination mechanism that rests on and extends the active inference framework for prediction-based perception, action, and decision-making \cite{Clark:2013jo,Adams:2012gn,Friston:2010ez}. 
This framework is based on embodied cognitive processing principles. We adopt it to model how agents acting in a situated task environment are influenced by what other agents are doing and intend to do, while only implicitly reasoning about the agents' on-the-fly coordination. 
Similar to related work, we develop and test this model in a simulated collaborative task domain 
inspired by the \textit{Overcooked}\cite{Overcooked} computer game. 
In this domain, multiple agents need to collaborate in preparing as many ordered meals as possible in a given kitchen environment and within a fixed amount of time.

With the proposed model, we aim to make two contributions. 
First, we present a \emph{hierarchical predictive processing}-based model for perception and action. This model is able to generate responsive, yet goal-directed behavior in an efficient and robust way. Crucially, it solves the planning problem without any explicit reasoning about or representation of the task structure (e.g., how exactly a specific meal needs to be prepared).
This is achieved by integrating bottom-up information, e.g., about current affordances for actions in the environment, with top-down information about the agent's goals. The integration yields the agent's current intentions and influences its actions and perception.
Second, we propose an implicit coordination strategy, called \emph{belief resonance}, that rests on integrating beliefs about another agent's intentions or goals, with beliefs about one's own actions and intentions. 
The degree to which another agent's intention influences one's own beliefs is controlled by a so-called \textit{susceptibility parameter} (SP). 
Using this parameter, different roles (e.g., leader or follower) can be realized by controlling how susceptible agents are to the mental states of another agent. In order to infer another's mental state efficiently, we integrate the active inference framework with a light-weight mentalizing approach based on a model of \textit{Bayesian Theory of Mind} (BToM) \cite{baker:2011bayesian}.

Framed within this model, we want to investigate to what extent the combination of these basic, neuro-cognitively inspired \cite{Friston:2015jr,Kilner:2007jq} 
mechanisms facilitates the emergence of efficient, real-time collaborative behavior. 
Clearly, this approach does not account for planning optimal actions or dividing labor into complementary sub-tasks, which requires complex representations and planning processes. 
Instead, we seek to explore what kind of collaborative behavior can emerge that is not in the repertoire of individual agents. To be more exact, agents are not explicitly reasoning about their collaborative task but are coupled through situated perception and action in a shared environment, along with a minimal social-cognitive ability for inferring and adopting others' intentions. 
The resulting implicit coordination processes impose lower computational requirements and are thus applicable for real-time interaction. 

In the remainder of this paper we first discuss related work and theoretical backgrounds. We will then present a \textit{hierarchical active inference for collaborative agents} (HAICA) model for agents engaged in situated collaborative tasks. As a foundation, we will initially focus on describing the computational model for belief updates across different layers, and then turn to the implementation of belief resonance in the model. 
Afterwards, we present the collaborative cooking task domain and describe how the model is applied to it. 
This also includes an explanation of an efficient ToM approach that we propose to use in HAICA. Furthermore, we present simulation studies with two agents acting in different environments in the Overcooked domain. 
We report results from evaluating the model while varying the SP values for each of the two agents in order to evaluate the effects of varying degrees of belief resonance on implicit coordination. 
We also investigate the effectiveness of belief resonance in asymmetric scenarios where one agent has more task knowledge than the other. Finally, we compare our results with current state of the art models both from reinforcement learning (RL) as well as probabilistic Bayesian models based on inverse planning.

\section{Related Work and Background} 

Developing models and systems for multi-agent collaboration has received increasing attention in fields such as autonomous systems, human-robot interaction, multi-agent systems, as well as Psychology and Cognitive Science. 
Work in Computer Science and A.I. has focused mostly on formalizing the coordination problem and finding ``optimal'' solutions through multi-agent planning \cite{torreno2017cooperative}.
Usually, agents are assumed to coordinate in advance, either explicitly or implicitly by sharing common communication protocols and/or planning algorithms. 
The strongest form of such prior coordination would involve a centralized control in charge of finding an optimal plan for the whole team \cite{dimarogonas2009event}.

Others have looked at multi-agent scenarios that do not allow for a full prior coordination or centralized control, but instead require agents to adapt to each other individually and dynamically, and even establish a form of ``ad-hoc teamwork'' \cite{barrett2017making,stone:2010ad,chen:2019atsis}. 
This problem of on-the-fly coordination has recently become prominent in collaborative robots and service robots \cite{mutlu2013coordination} and has been approached mostly for attention and motor-level coordination of physical tasks between humans and robots, e.g., when handing over objects or jointly carrying them \cite{mutlu2013coordination}. 
The underlying models of on-the-fly coordination emphasize the role of the agents' co-presence in a shared environment and their coupling through perception-action or force-feedback loops. 

On a related note, research in Psychology has explored how humans coordinate (both spatially and temporally) during joint action by means of perceiving, simulating and predicting others \cite{vesper:2017joint,nikolaidis:2017human}.
This work has provided substantial evidence that people rely on different mechanisms to coordinate their joint actions. 
These mechanisms do not only involve explicit planning but also lower-level sensorimotor coordination that yields an emergence of coordinated behavior \cite{knoblich2011psychological}.

A number of computational approaches to model key abilities for collaborating with others have been proposed. 
Examples include joint attention \cite{huang2011effects}, 
goal recognition \cite{vered2016online,masters2017cost}, 
online planning \cite{wu2011online}, 
or collaborative discourse \cite{rich2010collaborative}.
Most importantly for our work, significant advancements have been made in developing computational models of interpreting an  action of another agent in terms of the intentions, beliefs, or emotional states that may have caused said action. 
This so-called mind-reading, or Theory of Mind (ToM), requires the observer to have a good model of the acting agent. Commonly one assumes a naive rational agent that selects actions which maximize a subjective utility \cite{baker2009action,jara2020naive}. 
Earlier work defined such models in terms of the BDI framework \cite{bosse2011recursive}, while later work shifted to probabilistic models \cite{pynadath2011modeling} or Bayesian models based on inverse planning \cite{baker2009action,baker:2017rational,jern:2017people}. 
In the context of ToM, these inverse planning models have also been termed Bayesian Theory of Mind (BToM) \cite{baker:2011bayesian}. Such models have been able to fit empirical data and thus have great conceptual and theoretical merits, as they offer computational accounts of how humans reason and act in the social domain. 
However, the standard BToM framework requires a complete generative model for all actions, which is costly to acquire, formulate, and compute \cite{blokpoel2013computational}. Full BToM is thus too complex for on-the-fly coordination in real-time interaction (see Sect.~\ref{sec:results} for runtime measurements) and can only be applied in approximated schemes \cite{poeppel:2018satisficing}. 
Recently, work has also started to explore the use of deep learning to recognize different kinds of synthetic agents and to predict their future behavior \cite{rabinowitz2018machine}.

Only a few approaches have started to look at how artificial agents can be endowed with capabilities such as ToM to enable or improve continuous multi-agent collaboration. Clearly, implementations are always highly specific and attuned to the particular scenario they aim to model. Comparability of approaches thus is a major concern. 
One multi-agent collaboration scenario that has been adopted lately in various studies is from the Overcooked 
computer game \cite{Overcooked}, in which multiple agents need to work together to prepare ordered meals in a specific kitchen layout in a limited amount of time. In this task, agents need to coordinate on-the-fly on high-level sub-tasks in order to improve team performance. 

The Overcooked domain has been previously tackled using Deep Reinforcement Learning to train a collaborative policy from minimal domain knowledge \cite{carroll2019utility,song:2019diversity}. 
A BToM approach has also been used in a simplified version of the task \cite{wu2021too}.
In the latter approach, Wu et al. propose Bayesian Delegation, a Bayesian model that enables (joint) task inference using knowledge about sub-tasks with pre- and post-conditions to determine partial orderings of different recipes.
Each ``agent simulates a fictitious centralized planner that controls the actions of all agents working together on the same sub‐task'' (\cite[p. 422]{wu2021too}) in order to determine the best action to take for the agent. That is, the approach is to search for the most suitable collaborative action in order to complete the task in as few steps as possible.

In the present work, we ask how successful collaboration can emerge based on cognitively simpler means, without explicitly representing and solving an inherently complex multi-agent decision or planning problem. We use the Overcooked domain in the setting presented in \cite{carroll2019utility} and additionally evaluate our model in the version discussed in \cite{wu2021too} (Sect.~\ref{sec:results}). 
Our approach is related to the one presented in \cite{wu2021too} as we also employ a probabilistic BToM model to identify and adapt to another agent's underlying goals on-the-fly. 
We also provide domain knowledge in the form of possible (action) intentions, such as fetching a certain item or preparing an ingredient, and their likelihoods given an observation of the environment.
However, our approach is different as we do not feed ToM hypotheses into explicit action planning with pre- and post-conditions of different intentions. 
Furthermore, our model does not require explicit specification of how different intentions need to be ordered to complete a recipe. 
Rather, we want to investigate how mentalizing can be integrated with a prediction-based perception-action loop in an individual agent, and whether this can bring about multi-agent coordination in real-time when such agents come to interact in a shared environment. 

We base our approach on a large body of recent work on predictive processing and active inference. The underlying assumption is that action recognition rests on principles of \textit{predictive processing} \cite{Clark:2013jo} where predictions about sensory stimuli are continuously formed and evaluated against incoming sensory input. An agent's systemic goal is to minimize so-called free energy, i.e., its uncertainty in terms of its surprise and divergence with respect to its predictions \cite{Rao:1999in,Friston:2012fwa}. For action control, predictive processing views action as a form of \textit{active inference} by which the environment is affected to reduce uncertainty about predictions that stem from beliefs about the world \cite{Adams:2012gn}. The generative process is inverted to predict next actions and, thus, attenuate prediction errors \cite{Friston:2010ez}. This mechanism has been discussed as a form of so-called \textit{affordance competition}, a possible mechanism for action selection \cite{Cisek:2007fq} based on possible goals achievable through that action.

A common approach to cope with the complexity of sequential decision-making is to employ hierarchical models that make use of a domain's inherent structure \cite{herbortAdvantagesHierarchicalAnticipatory}.
However, relatively little work has tried to apply this approach to inter-agent collaboration. Friston et al.~\cite{Friston:2015jr} proposed to tackle the problem of inferring the intention behind communicative behavior by coupling two predictive processing-based models. 
Brandi et al.~\cite{Brandi:2019cd} proposed a predictive processing-based model to study interaction in terms of the participant's \textit{social agency}, i.e., their ability to predict a partner's contributions to an interaction. This approach, however, has not been applied to collaboration in real-time situated problem-solving.

The hierarchical model presented here builds on the previously proposed \textit{Hierarchical Predictive Belief Update} (HPBU) 
\cite{Kahl:2018ki}. 
HPBU was modeled after the assumed cortical micro-circuitry for predictive coding \cite{Bastos:2012bd} and is based on principles of active inference. It combines empirical Bayesian updates within a layer and linear dynamic updates (in form of a Kalman filter) between layers of its processing hierarchy. HPBU was developed as a generative model that learns, predicts and produces handwritten digits in an unsupervised way using hierarchical abstractions spanning from low-level movements to motor schemas corresponding to digits. This was used as a sensorimotor basis for a model of self-other distinction \cite{Kahl:2018ki} and was later extended to the social domain with additional levels 
that form a simplified mentalizing system to perform belief coordination in multi-agent scenarios \cite{kahlSocialMotoricsPredictive2020}.
In the present work we employ HPBU in more complex collaboration tasks and extend it with a mechanism for on-the-fly coordination as described next.


\section{Hierarchical Active Inference for Collaborating Agents (HAICA)}
In this section we present a model for \textit{hierarchical active inference for collaborative agents} (HAICA) as one building block for 
collaborative agents that are supposed to 
coordinate efficiently and on-the-fly with others co-situated in a joint task. 
We will first introduce the general principles underlying HAICA, before describing how it is implemented and applied to the Overcooked domain in Sect.~\ref{sec:overcookedInstantiation}.

\begin{figure}
    \centering
    \includegraphics[width=\linewidth]{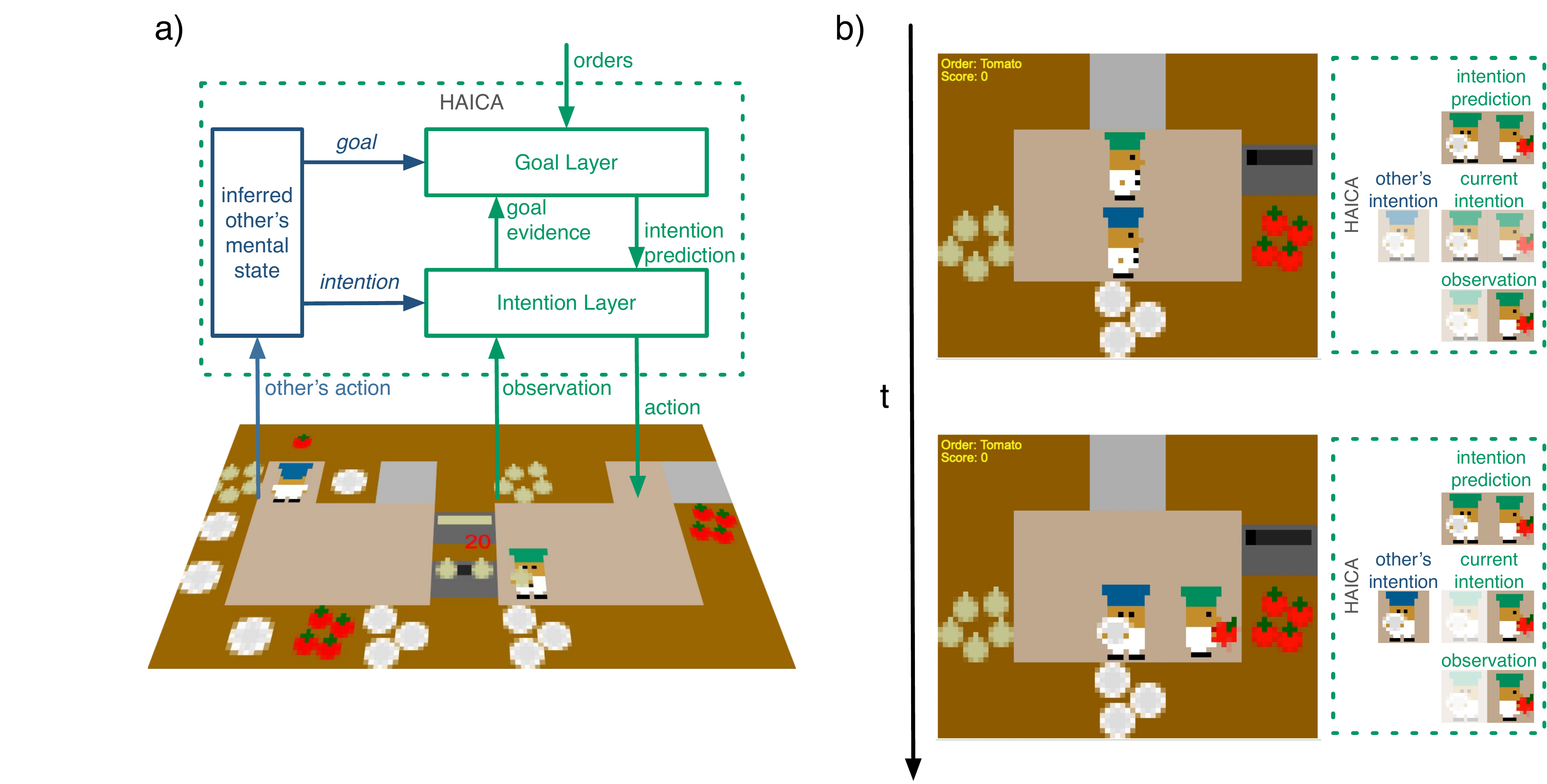}
    \caption{(a) Overview of the proposed model of hierarchical active inference for collaborative agents (HAICA) in the Overcooked domain, depicted for the green agent.
    Two predictive layers encode the agent's beliefs about its own intentions and goals. They are updated based on top-down (\emph{predictions}) and bottom-up (\emph{evidence}) information originating internally (in the agent) or externally (from the environment). 
    An additional component (for mentalizing or ToM; blue box) is included to infer the other agent's intentions or goals from its behavior. This information is integrated as additional beliefs in what we call belief resonance at the respective levels (see Sect.~\ref{sec:ToMIntegration}). 
    (b) Illustration of the integration process at the intention layer of the green agent. 
    Initially, the agent is not certain about its own intention.
    Given an order for \emph{Tomato} soup the goal layer predicts corresponding intentions for that order, e.g., picking up a tomato or plate. 
    At the same time, a belief about the other (blue) agent's intention is being inferred (getting a plate).
    Belief resonance would tilt the green agent towards the other's intention, i.e., getting a plate with increasing certainty.
    However, bottom-up evidence about the environment validates possible intentions resulting in the agent choosing to pick up a tomato since its path towards the plates is blocked.}
    \label{fig:hierarchy}
\end{figure}

The overall structure of the model is shown in Fig.~\ref{fig:hierarchy}(a), instantiated for the green agent. 
Every agent is equipped with a hierarchical model for behavior perception and generation based on predictive processing. 
The model governs the agent's behavior through generative processes working at different
levels of abstraction of the task representation. 
In this paper, we employ two layers that are geared towards the Overcooked domain, namely a \emph{goal} and an \emph{intention} layer.
The goal layer determines an agent's own desire with respect to which order to work on (e.g., a tomato soup), while the intention layer represents high-level actions that can be taken in the kitchen (e.g., picking up a specific item). 
Importantly, the layers represent the agent's beliefs about its \textit{own} current goals and intentions, respectively. These beliefs are subject to continuous, prediction-based update processes that take into account both top-down (\emph{predictions}) and bottom-up (\emph{evidence}) information. 
These information sources can be seen as proposed beliefs, e.g., concerning the agent's current intention, which are integrated to determine the new state of the respective layer (as described formally in Sect.~\ref{sec:beliefUpdate}). 
Crucially, this integration process is identical at each layer. 
The only differences are their domain and the source of the incoming information.
As shown in Fig.~\ref{fig:hierarchy}(a), for this domain the goal layer is concerned with the perceived orders (top-down predictions) as well as the currently available intentions (bottom-up evidence). The agent's current goal, in turn, proposes suitable intentions top-down to the intention layer, which integrates these proposed intentions with current beliefs about perceived action affordances in the environment (bottom-up). Naturally, these action affordances (as in pots, dishes and vegetables in a kitchen) are always also influenced by the other agent's actions.  
At any point in time, the agent selects an action in accordance with its most likely intention. 

The model's hierarchy (green solid boxes in Fig.~\ref{fig:hierarchy}(a)) is complemented with a mentalizing component (blue box) to enrich the agent's perception-action loop with additional information about the other agent that cannot be readily observed. 
For the integration of these components we propose a mechanism called \emph{belief resonance}, which allows 
the belief update of the agent's internal state to be affected 
by the inferred beliefs about another agent's intentions and goals. 
The idea is that -- similar to assumed resonance processes in the human sensorimotor system -- agents can in that way align at the level of intentions and goals, which may implicitly lead to better coordinated actions.
Note that HAICA does not make any assumptions about how the mentalizing component works, it only requires the eventually inferred beliefs about the other agent's mental states. 
Fig.~\ref{fig:hierarchy}(b) illustrates the complete integration process for the intention layer.
Furthermore, HAICA does not specify explicitly when or in what order intentions are to be considered. Goal directed and coordinated behavior instead emerges solely from the hierarchical belief update. 
Top-down information, here in the form of intention predictions originating from the agent's current goals, specifies the intentions that are required to achieve that goal, e.g., a specific meal order, without imposing an ordering of these steps. Next intentions are selected only by validating what should be done in general against what actions are currently afforded by the environment (given in the form of bottom-up evidence). 

Next, we describe the computational mechanism for the belief update process at each layer in the hierarchical model (Sect.~\ref{sec:beliefUpdate}), and then present how \emph{belief resonance} is formalized in this model (Sect.~\ref{sec:ToMIntegration}). The domain-dependent implementation of mentalizing is explained in Sect.~\ref{sec:ToM} after we have introduced the Overcooked domain in detail.

\subsection{Integrating top-down and bottom-up information \label{sec:beliefUpdate}}

Each layer $L_j$ represents a certain abstract mental state related to the task domain, e.g., the goal layer represents the different goals or desires the agent may have.
A layer $j$ defines a probability distribution $P_t(L_j) := P_t(s_j) ~~\forall s_j \in L_j$ across its domain $L_j$ to capture the degrees of belief the agent has regarding its own state. These distributions are updated every time $t$ the agent perceives new observations. 
More specifically, as shown in Fig.~\ref{fig:intention_belief_update}, each layer is updated through probabilistic information from the layer above ($L_{j+}$), in the form of \emph{top-down} predictive influences, as well as from the layer below ($L_{j-}$), as \emph{bottom-up} evidence. 
This information takes the form of assumptions the adjacent layers hold about what layer $j$'s state should look like from their current perspective. 
In the case of outer layers in the hierarchy, the layers receive this kind of information directly from the environment. 
Both top-down and bottom-up information are first integrated with the prior ($P_{t-1}(L_j)$), i.e., the layer's previous belief state.

The integration is performed using a Kalman filter, resulting in separate top-down predictions about the layer's new state from the next higher layer $L_{j+}$ as $P_t^{j+}(L_j)$ and bottom-up evidence from the next lower layer $L_{j-}$ as $P_t^{j-}(L_j)$: 

\begin{figure}
    \centering
    \includegraphics[width=0.65\textwidth]{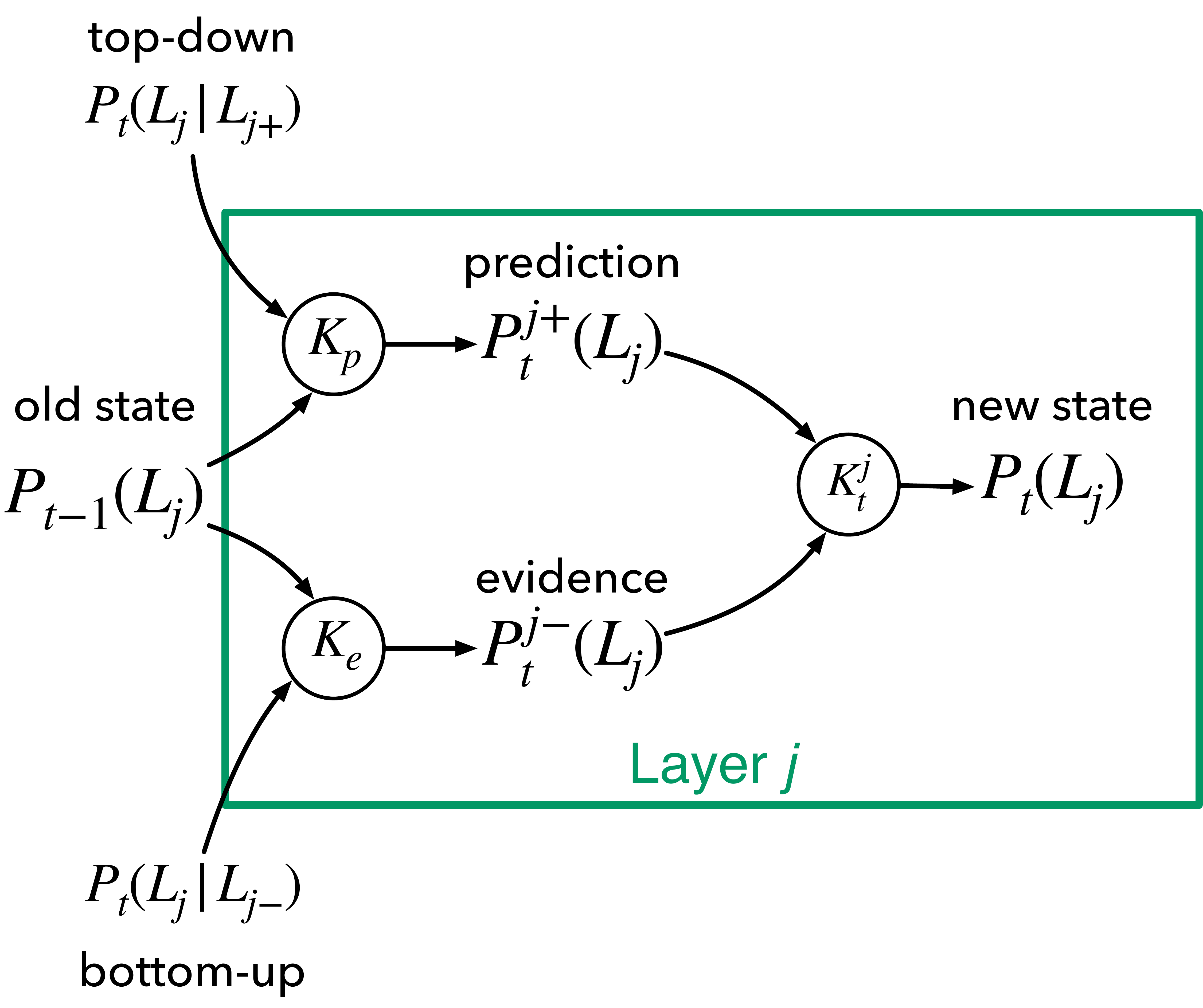}
    \caption{
    \label{fig:intention_belief_update} Realization of the belief update within a predictive layer. Each circle represents a Kalman filter integration with the respective Kalman gains. 
    Top-down $P_t(L_j|L_{j+})$ and bottom-up $P_t(L_j|L_{j-})$ likelihoods are first integrated with the empirical prior $P_{t-1}(L_j) := P_{t-1}(s_j) ~~\forall s_j \in L_j$ to produce the prediction $P_t^{j+}(L_j)$ and evidence $P_t^{j-}(L_j)$.
    $K_p$ and $K_e$ are fixed meta-parameters but $K_t^j$ is computed by based on the previously computes predictions and evidence.
    The final posterior $P_t(L_j)$ is computed by integrating the top-down prediction with the bottom-up evidence.}
\end{figure}


\begin{align}
P_t^{j+}(L_j) := P_t^{j+}(s_{j}) &= P_{t-1}(s_j) + K_p[P_t(s_j|L_{j+}) - P_{t-1}(s_j)] ~~\forall s_j \in L_j\\
P_t^{j-}(L_j) := P_t^{j-}(s_{j}) &= P_{t-1}(s_j) + K_e[P_t(s_j|L_{j-}) - P_{t-1}(s_j)] ~~\forall s_j \in L_j
\end{align} 

Here, $K_p$ and $K_e$ are the Kalman gains for the prediction and evidence, respectively. 
These are meta-parameters that can increase the importance of specific information if desired. 
Usually, they can both be set to $0.5$, thus weighing incoming and prior information equally. 
$P_t(L_j|L_{j+})$ and $P_t(L_j|L_{j-})$ are the top-down predictions and bottom-up evidence, respectively. We compute these distributions by treating the outer layers as soft evidence:
\begin{align}
P_t(L_j|L_{j+}) := P_t(s_j|L_{j+}) = \sum_{s^{j+} \in L_{j+}} P(s_j|s^{j+}) P_t(s^{j+}) ~~\forall s_j \in L_j
\end{align}

$P(s_j|s^{j+})$ is the likelihood function for a specific state $s_j$ in $L_j$, given a specific state $s^{j+}$ in $L_{j+}$. 
$P_t(L_j|L_{j-})$ is computed analogously but with $P_{t-1}(s^{j-})$ as we update layers in decreasing order.

Generally, a Kalman filter is a linear dynamic update that provides the optimal state estimate of an underlying (noisy) signal \cite{kalmanNewResultsLinear1961}.
Since in our case the information sources (a layer's top-down prediction, bottom-up evidence and others' inferred state) are inherently noisy, we chose the Kalman filter to estimate the true state based on these noisy proposals. 
Kalman filters have previously been proposed as a model for optimal state estimation in sensorimotor integration \cite{Wolpert:1995hc}, for end-stopping and other extra-classical receptive field effects \cite{Rao:1999in}, and they have been used as a forward model for internal state estimation in optimal feedback control \cite{Todorov:2002bf}.

At this point we have two different distributions concerning the layer's state, the evidence $P^{j-}_t(L_j)$ and the prediction $P^{j+}_t(L_j)$.
The final belief update mechanism is based on previous work \cite{Kahl:2018ki}
and combines empirical Bayesian updates with an additional Kalman filter that uses a layer-specific Kalman gain $K_t^j$.
$K_t^j$ is computed based on a ``free energy'' $F_t^j$ and a ``precision'' $\pi_t^j$, i.e., the inverse variance of the prediction error ($\textit{PE}_t$)). 
The integrated posterior $P_t^j$ is then approximated as follows:

\begin{align}
    P_t(L_j) := P_t(s_j) &= P_t^{+}(s_j) + K_t^j (P_t^{j-}(s_j) - P_t^{j+}(s_j)) ~~\forall s_j \in L_j \label{eq:belief_update} \\
    K_t^j &= \frac{F_t^j}{F_t^j + \pi_t^j} \label{eq:kalman_gain} \\
    \pi_t^j &= \text{ln}\frac{1}{\text{var}(\textit{PE})} \label{eq:precision} \\
    \textit{PE}_t(L_j) &= P_t^{j-}(s_j) - P_t^{j+}(s_j) ~~\forall s_j \in L_j \label{eq:prediction_error}\\
    F_t^j &= H(P_t^{j+}(L_j)) + D_\text{KL}(P_t^{j+}(L_j)||P_t^{j-}(L_j))  \label{eq:free_energy}
\end{align} 

Eq.~\ref{eq:free_energy} consists of the entropy $H(P_t^{j+}(L_j))$ of the prediction distribution and the cross-entropy $D_\text{KL}(P_t^{j+}(L_j)||P_t^{j-}(L_j))$ between the prediction and the evidence. 
This way, the integrated posterior $P_t(L_j)$ results from the precision-weighted bottom-up and top-down information.
The resulting posterior is then used to compute the likelihoods for the next higher (as evidence) and next lower (as prediction) levels to inform their updates. 
This integration of information is sensitive to the uncertainty at every level, as the used Kalman gain depends on the precision of the incoming information. 

During interaction with the environment, any information inferred at a layer has to be integrated with higher-layer predictions so that the resulting behavior generation does not discount important sensory evidence or disregard helpful predictions from more abstract reasoning or richer context. 
To that end, both bottom-up and top-down information is weighted and integrated in a way that is sensitive to the uncertainty in the available information. 
Generally, making correct predictions about hidden causes under uncertainty is tricky and the influence of prediction errors on prior predictions has to be balanced carefully. 
This balancing act should depend on the uncertainty itself and is accounted for by \emph{precision weighting} of the prior predictions \cite{Friston:2009iz}. 

\subsection{Belief Resonance \label{sec:ToMIntegration}}

In order for an agent to make use of the inferred mental state of another agent $Q_t(L_j) := Q_t(s_j) ~~\forall s_j \in L_j$, this third source of information is integrated through a mechanism of \emph{belief resonance}. 
The core idea is to make the agent's self-beliefs about its own intentions and goals (which are affecting its actions by way of active inference) affected by its current beliefs about the intentions and goals of the interaction partner. That is, we assume that the agent has inferred a distribution $Q_t(L_j)$ about the mental state of another agent. 
While HAICA does not make any assumptions here about how this mental state has been inferred, we do note that on-the-fly coordination imposes real-time requirements. 
Classical inverse-planning or other full BToM adaptations can quickly become computationally too demanding without the use of strong priors \cite{baker2009action}. 
Since we are focusing on efficient models for on-the-fly coordination here, we use a ``satisficing'' (i.e. satisfying and sufficing) approach \cite{poeppel:2018satisficing} with a BToM model adapted to the Overcooked domain. This approach will be presented in detail in Sect.~\ref{sec:ToM}.

In order to integrate the inferred distribution $Q_t(L_j)$, through what we call here \emph{belief resonance}, we expand the belief update scheme in Eq.~\ref{eq:belief_update} to model an influence of inferred others' beliefs on one's own action-inducing beliefs at each layer. This results in the updated model shown in Fig.~\ref{fig:belief_resonance} and defined in Eq.~\ref{eq:tom_belief_update_0} below.

\begin{figure}
    \centering
    \includegraphics[width=0.7\textwidth]{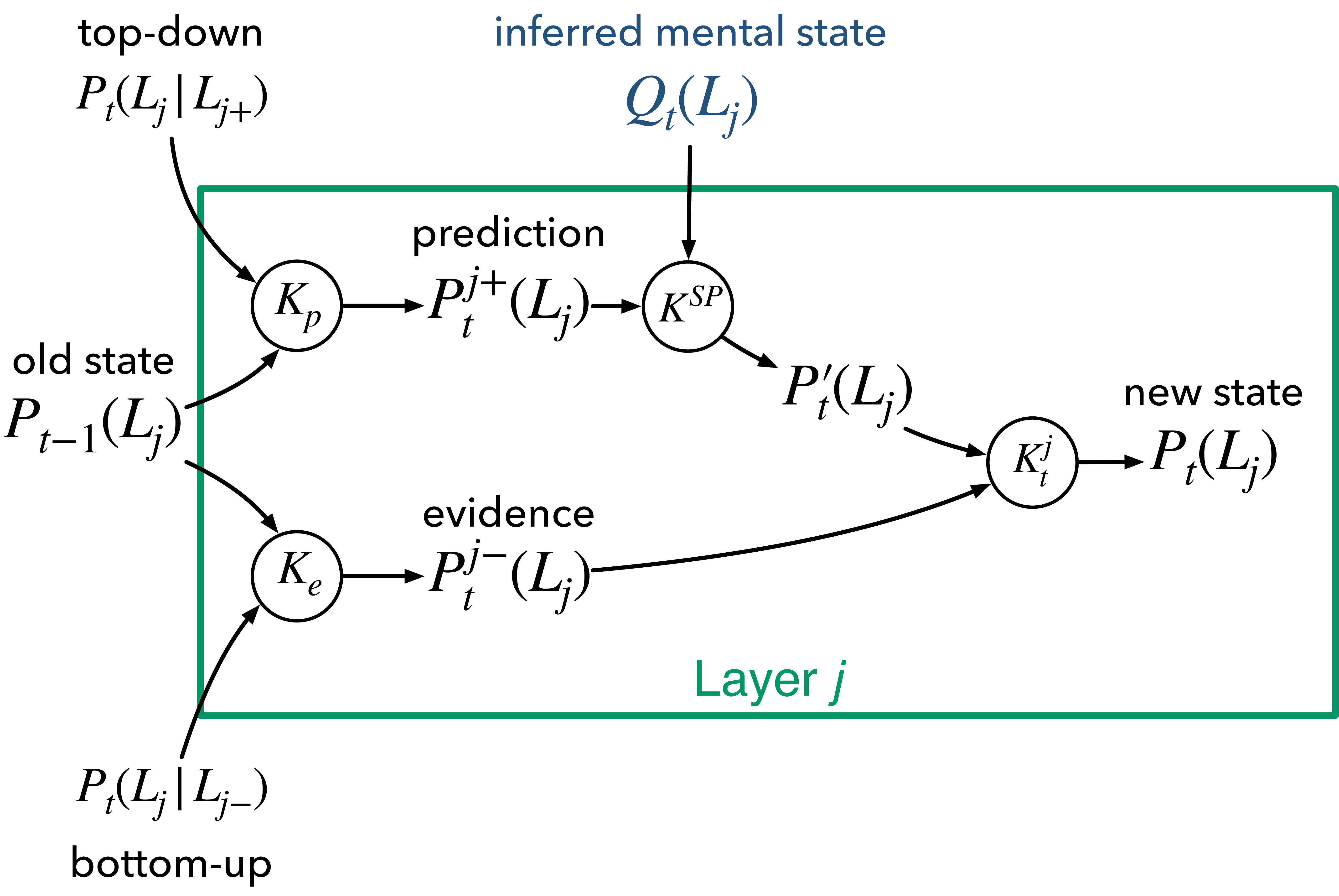}
    \caption{\label{fig:belief_resonance}
    Extension of the belief update model in Fig.~\ref{fig:intention_belief_update} with belief resonance. 
    The top-down prediction $P_t^{j+}(L_j)$ is first updated using the inferred mental state $Q_t(L_j) := Q_t(s_j) ~~\forall s_j \in L_j$ using a Kalman filter with susceptibility parameter $K^{SP}$ as gain, before updating and validating this with the agent's own bottom-up evidence $P_t^{j-}(L_j)$ for the final posterior $P_t(L_j)$.
    } 
\end{figure}

Following the tenet of developing a minimal model, we chose to treat the inferred beliefs $Q_t(L_j)$ as another source of information in addition to bottom-up or top-down information. 
That is, the integration scheme does not differentiate, at the level of internal beliefs, between own actions and others' actions. This is similar to the resonance processes assumed for motor action and perception in the human sensorimotor system \cite{schutz:2007perceptual}, but now applied to the level of abstract beliefs about action intentions and goals. We conjecture that this may be sufficient to implicitly lead to inter-agent coordination. 

Crucially, belief resonance is contingent on the (un-)certainty inherent in the respective beliefs. 
Further, the adoption of the inferred (ToM) beliefs is modulated by a \textit{susceptibility parameter} (SP) that we realize using another Kalman gain $K^\text{SP}$. 
This gain factor controls the influence of the inferred beliefs on the agent's posterior belief distribution at each layer.
While we set a fixed SP for the agents in this paper, in general, we do not assume this gain factor to be fixed or identical 
for every interaction partner or situation. 
Rather, it can be used to model roles, personality traits, or may be updated continuously throughout the interaction based on an evaluation of the interaction partner.

Mathematically, \textit{belief resonance} consists of a two-step belief update that first integrates the agent's inferred evidence $Q_t(L_j)$ for layer $j$ with the agent's predictions $P_t^{j+}(L_j)$ into an intermediate posterior $P_t^{\prime}(L_j)$.
$P_t^{\prime}(L_j)$ is then updated and validated with the agent's own evidence $P_t^{j-}(L_j)$ for the final posterior $P_t(L_j)$.
This is an extension of the dynamic belief update previously proposed in Eq.~\ref{eq:belief_update}:

\begin{align}
P_t^{\prime}(L_j) &:= P^{\prime}(s_j) = P_t^{j+}(s_j) + K^\text{SP} (Q_t(s_j) - P_t^{j+}(s_j)) ~~\forall s_j \in L_j \label{eq:tom_belief_update_0} \\
P_t(L_j) &:= P_t(s_j) = P_t^{\prime}(s_j) + K_t^j (P_t^{\prime}(s_j) - P_t^{j-}(s_j))  ~~\forall s_j \in L_j
\label{eq:tom_belief_update} 
\end{align} 

Note that Eq.~\ref{eq:tom_belief_update_0} only takes into account the top-down information for layer $j$ in the form of $P_t^{j+}(L_j)$ as well as the information inferred for the other agent at the same layer $Q_t(L_j)$. 
Besides using the special Kalman gain $K^\text{SP}$, it is functionally equivalent to the general belief update rule applied to the evidence from inferred beliefs.
Belief resonance can thus be understood as updating the agent's internal beliefs from both internal and (social) external sources of evidence.

In this way, the SP influences how strongly an agent is susceptible to the inferred mental states of the other agent. 
Higher values will cause a stronger resonance as $Q_t(L_j)$ will have a stronger influence on $P_t^{\prime}(L_j)$.
This could result in a shift in the agent's own beliefs, at least if the inferred beliefs are fairly certain, i.e., few beliefs have high probabilities. 
In contrast, low SP values will reduce the impact of the inferred mental states.
Note, that Eq.~\ref{eq:tom_belief_update_0} will simplify to $P_t^{\prime} = P_t^{j+}(L_j)$ if SP$=0$, resulting in the same model presented in Fig.~\ref{fig:intention_belief_update} or Eq.~\ref{eq:belief_update}.
At the same time, the agent's own bottom-up information is taken into account to validate the resonating beliefs for any given situation. 
Below, we will briefly show the effect of integrating the belief about the other agent after the bottom-up information, to demonstrate the importance of this validation.

\section{Applying HAICA to the Overcooked domain}
In order to evaluate the proposed HAICA model, we have applied it to collaboration tasks in the Overcooked domain. In this section, we introduce this Overcooked domain in detail and describe how the layers of the model with their respective likelihood functions are realized in this particular domain. 
Afterwards, we present an adapted BToM framework for efficient mentalizing as required for belief resonance. Finally, we describe how specific intentions of the agents are turned into actual behavior.

\subsection{The Overcooked domain} 
The domain is inspired by the video game Overcooked \cite{Overcooked}.
In this game, multiple agents work together in different kitchen layouts in a 2D grid-world, to prepare as many ordered meals as possible in a limited amount of time. 
In general, a team is rewarded for every order it completes, with a new order replacing the fulfilled one. 
Different orders may give a different amount of points depending on the complexity of the order.

Within the environment, agents can move in the four cardinal directions to the next square (if it is free) and turn towards that direction. 
Furthermore, agents can pick up objects like ingredients (onions,  tomatoes or lettuce) or plates by interacting with these objects. 
Held objects can be dropped onto a free counter in a similar manner. 
The environment may further contain special objects, such as a cutting board or a cooking pot that an agent can interact with in order to ``prepare'' ingredients for the different orders.

For this work, we implemented the Overcooked environment similarly to Carroll et al. \cite{carroll2019utility}:
In this implementation, agents need to explicitly perform an ``interact'' action in order to pick up or drop an item and interact with the cooking pot or delivery tile. 
Agents need to prepare different kinds of soups (\textit{Onion} or \textit{Tomato} soup) using a cooking pot. The environment always provides two orders 
for the agents to work on.
Each soup is cooked by putting three ingredients of the respective type (onion or tomato) into the cooking pot. 
The ingredients and plates can be taken from ``dispensers'' which provide an unlimited amount of these objects.
Once a pot contains all the ingredients for a recipe it will start cooking. Each different flavor of soup requires a different but fixed amount of time steps to cook.
The cooked soup can then be taken out by an agent by interacting with the pot while holding a plate.

Since the soup requires time to cook, agents should use that time to work on another order in the meantime to increase efficiency. 
This interaction favors coordination on a higher level, as agents should start preparing the next order when they realize that all ingredients for the first order are already being prepared by another agent. 

\begin{figure*}
    \includegraphics[width=\textwidth]{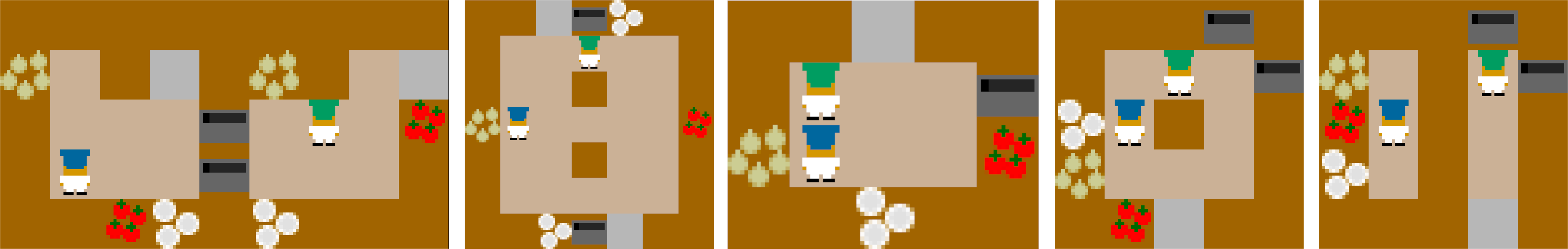}
    \caption{Five different kitchen layouts in the Overcooked domain that increasingly require multi-agent coordination (from left to right): \emph{asymmetric, spacey, cramped, ring} and \emph{forced}.
    Visual assets adapted from \cite{carroll2019utility}.}
    \label{fig:kitchen_layouts}
\end{figure*}

We consider the five different kitchen layouts shown in Fig.~\ref{fig:kitchen_layouts}: 
\emph{asymmetric, spacey, crowded, ring} and \emph{forced}, from left to right. 
Most of these layouts were also applied in the study by Carroll et al. \cite{carroll2019utility} and are well-suited to evaluate collaboration as they pose different demands:
The first layout (\emph{asymmetric}) has the agents physically separated but sharing two pots in the middle. 
As there is no interference, hardly any coordination is required and either agent could solve the task by itself. 
The second layout (\emph{spacey}) is similar in that the agents could solve the problem individually. 
However, a valid collaboration strategy may involve splitting up the orders among the agents.
The remaining kitchen layouts require increasing amounts of coordination in order for the agents to be able to solve the task. 
Agents need to avoid blocking the paths of the other agent in the \emph{cramped} layout and even more so in the \emph{ring} layout. Finally, the \emph{forced} layout requires the two agents to collaborate as the ingredients and the pots are physically separated. Each agent only has access to either the ingredients or the pots. 

\subsection{Implementation of HAICA in Overcooked} \label{sec:overcookedInstantiation}

As shown in Fig.~\ref{fig:hierarchy} and described above, we realize two layers of the HAICA model in the Overcooked domain. The \emph{goal} layer represents the agent's desire at a given time and holds a distribution across the different orders an agent may work on (\textit{Onion} or \textit{Tomato} soup). This layer receives its top-down influence directly from the orders, with a likelihood that represents the order distribution, e.g., 50/50 if both Onion and Tomato soups are ordered. The second layer represents the \textit{intentions}, i.e., high level-actions an agent may carry out. All possible intentions and their parameters are listed in Table~\ref{tab:intentions}.

\begin{table*}
\centering
  \caption{Intentions of agents in the Overcooked domain. Parameter abbreviations stand for (O)nion, (T)omato, (D)ish and (S)oup.}
  \label{tab:intentions}
  \begin{tabularx}{\textwidth}{ccX}
    \hline\noalign{\smallskip}
    Intention &  Parameters & Meaning\\
    \noalign{\smallskip}\hline\noalign{\smallskip}
    get-item & O, T, D, S & Pick up the closest item \\
    drop-item & O, T, D, S & Drop item on the closest free counter \\
    interact-with-pot & Pot & Interact with the concrete pot \\
    deliver-soup & T, O & Deliver soup to serving tile \\
    hand-over & O, T, D, S & Drop item on the closest free counter accessible by both agents\\
    wait & - & Do nothing \\
  \noalign{\smallskip}\hline
\end{tabularx}
\end{table*}

The intentions can be seen as the agent's actionable capabilities within the environment. In order to be able to solve the task at all, an agent needs to be able to pick up and drop items and interact with the varying objects in the environment.
The ``hand-over'' intention may appear like hard-coding coordination. However, this intention only provides the agent with the capability to hand an object over.
As mentioned above, it is not explicitly specified when or how this intention is to be used in order to aid collaboration. Instead, HAICA will choose the intention when it becomes most likely according to the top-down predictions based on the agent's goal and the bottom-up affordances perceived in the environment. 

Each of these higher-level intentions requires a sequence of actions in the environment. How these intentions are turned into low-level actions is explained in Sect.~\ref{sec:intentionToAction}.

\subsubsection{Modeling domain-specific intention likelihoods}

For each of the intentions listed in Table~\ref{tab:intentions}, we defined functions to compute their likelihood given the current state of the environment. 
To show the robustness of the model and to ensure real-time capability, the likelihood functions used here are primarily affordance checks.
Specifically, most of these functions return either $0$ or $1$ which are then normalized over all intentions for the actual likelihood. 
A function returns $0$ in case an intention is not applicable in the current state, e.g., an agent cannot pick-up another item if it is already holding something in its hands. 
The only exception here is the likelihood for \emph{wait} which is simply a fixed value of $1/|intention|$ ($\approx 0.08$ with the instantiated intentions outlined above) before normalization, to indicate that the agent cannot currently do anything useful. 

More fine-grained control as well as better performance can be achieved by optimizing these likelihoods either through manual fine-tuning or learning from data. 
For example, while the agent can solve the task with the current implementation of \emph{interact-with-pot}, it would be advantageous to take the distance to the pots as well as the number of ingredients already in them into account. That way, the agent could favor a pot that it already started to fill before starting to fill a second pot.

For the evaluation reported below, we modeled this heuristic along with some slack, i.e., non-zero probability for delivering a soup that was not ordered and dropping an item that is currently usable with a pot or the serving tile. 
The ``hand-over'' likelihood is reduced if the other agent is currently holding something or if the agent could handle the held ingredient itself.
The likelihood functions in-between the two layers for both, top-down from \emph{goal} to \emph{intention} as well as bottom-up from \emph{intention} to \emph{goal}, are similarly normalized affordance checks. 
For instance, the likelihood for \emph{get-item(Onion)} given the goal \textit{Tomato} soup would be $0$ since \textit{Tomato} soup does not require an onion. 

Overall, the likelihoods implicitly encode the agent's recipe knowledge. 
The inter-layer likelihoods define which steps are required for a certain meal in general, e.g., picking up tomatoes or plates, interacting with a pot and delivering the soup, but not their ordering. The idea is that a goal-directed ordering of intentions emerges since the bottom-up likelihoods invalidate intentions that are not (yet) applicable. In the example illustrated in \ref{fig:hierarchy}(b) the intention of picking up a plate is invalidated due to the environmental constraints. 
By choosing these deterministic inter-layer likelihoods we assume that each agent knows the steps required for the different orders.

\subsubsection{Efficient inference of other agents' mental states \label{sec:ToM}}

The proposed mentalizing approach is based on the BToM framework, using a generative model to define how different mental states influence an agent's behavior.
Here, we consider the mental states corresponding to the two central layers that the hierarchical model assumes, namely possible goals (or orders) (G) and intentions (or abstract actions) (I) an agent may have. This results in the following generative model:

\begin{align}
P(a,o) &= \sum_{i \in I, g \in G} P(a|i,g,o)\cdot P(o|i,g) \cdot P(i|g) \cdot P(g)
\label{eq:general_tom_update}
\end{align}

where $P(a|i,g,o)$ defines the likelihood of an action $a$ given a certain intention $i$, a goal $g$ and current state observations $o$.
$P(o|i,g)$ models the likelihood for specific state observations and $P(i|g)$ models the likelihood of an intention given the specific goal. 

This model can be inverted using Bayes' rule:

\begin{align}
P(i|a,o) & \propto \sum_{g \in G} P(a|i,g,o) \cdot P(o|i,g) \cdot P(i|g) \cdot P(g) \\
P(g|a,o) & \propto \sum_{i \in I} P(a|i,g,o) \cdot P(o|i,g) \cdot P(i|g) \cdot P(g)
\label{eq:inferred_metal_states}
\end{align}

Generally, inferring the mental states G and I from this generative model would require us to evaluate all possible actions. 
Further, the likelihood function $P(a|i,g,o)$ itself may be difficult to compute depending on the scenario. 
As an example, Wu et al. \cite{wu2021too} follow the common approach to set the action likelihood proportional to the soft-max of the expected future reward $Q^*(s,a)$, modulated by a ``rationality'' parameter. 
This requires the agents to compute estimations of the expected rewards, which in itself can be computationally very demanding as can be seen in our runtime comparison below.

Since collaboration often requires timely reactions and using a complete generative model quickly becomes too costly \cite{poeppel:2018satisficing}, we opted for an approximate but ``satisficing'' solution. This is achieved by approximating the likelihood $P(a|i,g,o) = \alpha$ if $a = a^*$, with $a^*$ being the action the agent would take in the situation $o$ if it had intention $i$ and goal $g$. 
We can compute $a^*$, e.g., by utilizing the agent's own action selection capabilities. 
While the classical soft-max approach models an agent's action likelihoods to correlate with ``rational'' behavior (usually regulated by the ``rationality'' parameter), our simplification assumes the other agent should behave similar to oneself. 
We use $\alpha$ to model how similar we are assuming the other agent to be. Low values of $\alpha$ will lead to more random actions, different from what the agent would do itself. 

A general problem with probabilistic integration over time is that of disappearing probabilities. 
Integrating several very small likelihoods, e.g., for a particular intention, will result in probabilities approaching the numerical zero due to rounding errors. 
Counter-evidence in the form of strong likelihoods for the same intention will then hardly influence the posterior at that point.
This results in a system that would react very slowly to changes in the behavior of the other agent.
Since we require quick adaptation to the other agent in the Overcooked scenario, we employ a \textit{softmax} normalization after updating the inferred beliefs in conjunction with adding a fixed noise $\mu$ prior to normalization:

\begin{align}
P'(i|a,o) & \propto \exp{\lbrack \beta \cdot (P(i|a,o)+ \mu) \rbrack} \label{eq:normalized_inferences0} \\
P'(g|a,o) & \propto \exp{\lbrack \beta \cdot (P(g|a,o)+ \mu) \rbrack} 
\label{eq:normalized_inferences}
\end{align}

where $\beta$ controls the strength of the maximization. 
Small values for $\beta$ will result in pushing the distribution towards uniformity, while large values will reinforce the largest outcomes of the distribution. 
$\mu$ is used to ensure that all outcomes have relevant probability masses so that the system can react quickly to changing likelihoods. 
These updated posterior distributions concerning the intentions and goals are then used as $Q_t(L_j)$ depending on the layer in the belief resonance description in Sect.~\ref{sec:ToMIntegration}.
If layer $j$ is the intention layer, $P'(i|a,o)$ is used as the inferred belief, while we use $P'(g|a,o)$ in the goal layer.

The used approximation of $P(a|i,g,o)$ can be replaced with a handcrafted or learned model of the proper likelihood if required.
However, as shown below, even in its simple form it proves to be sufficiently reliable and thus demonstrates how simple ToM inferences can be achieved without a full generative model.
Further, since this approach only requires discrete action predictions, it can be combined with predictive processing models, such as the HAICA model employed here, as well as more traditional planners.

\subsubsection{Processing cycle and turning intentions into actions} \label{sec:intentionToAction}

Every time an agent receives new observations from the environment it first uses the ToM module to update its beliefs regarding the other agent's mental state, according to Eq.~\ref{eq:normalized_inferences0} and \ref{eq:normalized_inferences}.
The agent then computes the top-down and bottom-up beliefs at both the goal and intention layer, before integrating all information using Eq.~\ref{eq:tom_belief_update_0} and \ref{eq:tom_belief_update} to form the new posteriors. 
The goal layer receives the current orders as top-down influence, while the intention layer receives the different likelihoods based on the agent's current observations.
The agent's last posterior is used as prior, unless it received a reward in which case its mental state is reset in preparation for the next task.
Then, the maximum a posteriori (MAP) intention is picked as the agent's new active intention, i.e., the intention it is actually going to act upon. 
We regard this hierarchically informed and precision-weighted selection process of an active intention as a form of active inference. 


Setting the next active intention must be followed by planning the low-level actions to fulfill it. Wu et al. \cite{wu2021too} used bounded real‐time dynamic programming (BRTDP) to find an optimal policy for a selected sub-task.
We opted for a simpler approach in favor of greater computational efficiency.
Generally, intentions relate to interactions with certain objects in the environment. This means that in order to fulfill an intention, the agent usually needs to move to the target object, which could be an ingredient or a preparation object, e.g., the cooking pot. 
Given that the intention tells us what object we want to interact with, we can employ a best-first search algorithm like A* to find appropriate low-level actions (Left, Right, Up, Down, Interact or Wait) for a selected intention. 
Note that, while very simple and efficient, this approach to action planning does not take the other agent's likely future actions into account. As a result, one agent may block access to certain objects that another agent may need.

Due to the dynamic nature of the task, an agent may need to change its intention before achieving it.
One example would be that another agent has already fulfilled a certain step in the recipe or because the other agent blocks access to certain resources. 
Since the agent is updating its mental state every time it receives new observations, the model is able to cope with these challenges inherently and immediately. 
Still, one important addition we made is that of ``punishing aborted intentions''. 
Due to the dynamic nature of the model, when two agents each equipped with this model collaborate, they may end up switching back and forth between two alternating intentions, each of which invalidating the current intention of the other agent. 
Instead of employing an explicit memory in the form of a state-intention history, we opted to only check if the agent's current intention changed without achieving it.
If the previous intention was aborted, we punish it by reducing its prior probability for the next round of belief updates. 

\section{Results \label{sec:results}}

To study if and how the proposed HAICA model can enable agents to collaboratively solve joint tasks, we ran a series of simulations in order to investigate the following aspects. 
Firstly, we analyze if the HAICA model is able to solve the tasks at all, as it does not involve explicit task-planning or coordination. 
Secondly, we explore if and when belief resonance is beneficial for collaboration between two agents. To that end, each agent is equipped with the belief resonance model but with varying SP values. We expect to find layouts with higher demands for collaboration, such as \emph{ring} or \emph{forced}, to be solved best by teams with agents that are susceptible to one another's mental state. 
As an additional condition, we evaluate teams in which either both agents know the orders or only the first agent (hence requiring the second agent to infer which meals need to be prepared from the first agent's behavior). 
Such a scenario represents a leader-follower dynamic.
Thirdly, we evaluate the importance of validating resonated beliefs via bottom-up evidence. For that, we test an alternative order to integrate the different information sources, i.e., we swap $P_t^{j-}(L_j)$ with $Q_t(L_j)$ as well as $K_t$ with $K^{SP}$ in Eqs.~\ref{eq:tom_belief_update_0} and \ref{eq:tom_belief_update}. 
This corresponds to the agent integrating its own information sources first (top-down and bottom-up), before integrating the inferred belief about the other.
Finally, we directly compare the HAICA model with two state of the art solutions presented in \cite{carroll2019utility} and \cite{wu2021too}, respectively.
Although the proposed model is focused more on efficient on-the-fly collaboration, this comparison will allow us to put our results into a larger perspective with regard to the relation between optimality and efficiency.

All results reported in the following were obtained by averaging team rewards achieved in simulations with over 20 episodes of 400 time steps each.
New orders (\textit{Onion} or \textit{Tomato} soup) were generated at random, rewards were obtained after having prepared a meal and corresponded to the respective cooking times: 20 points for \textit{Onion} soup and 15 for \textit{Tomato} soup. 
Agents always started an episode at the same positions (see Fig.~\ref{fig:kitchen_layouts}), the SP values were assigned at random to the two agents to avoid influences of their starting points.
Table~\ref{tab:meta_parameters} lists the values for all used meta parameters. 

\begin{table*}
\centering
  \caption{Meta parameters used for the evaluation}
  \label{tab:meta_parameters}
  \begin{tabularx}{\textwidth}{ccX}
    \hline\noalign{\smallskip}
    Parameter &  Value & Explanation\\
    \noalign{\smallskip}\hline\noalign{\smallskip}
    $\alpha$ & 0.9 & Action likelihood of predicted action in ToM module \\
    $\beta$ & 2 & Softmax parameter for ToM module \\
    $\mu$ & 0.1 & Random noise for ToM module \\
    $K_e, K_p$ & 0.5 & Kalman gains for bottom-up and top-down belief integration \\
  \noalign{\smallskip}\hline
\end{tabularx}
\end{table*}

\begin{figure}
    \centering
    \includegraphics[width=\linewidth]{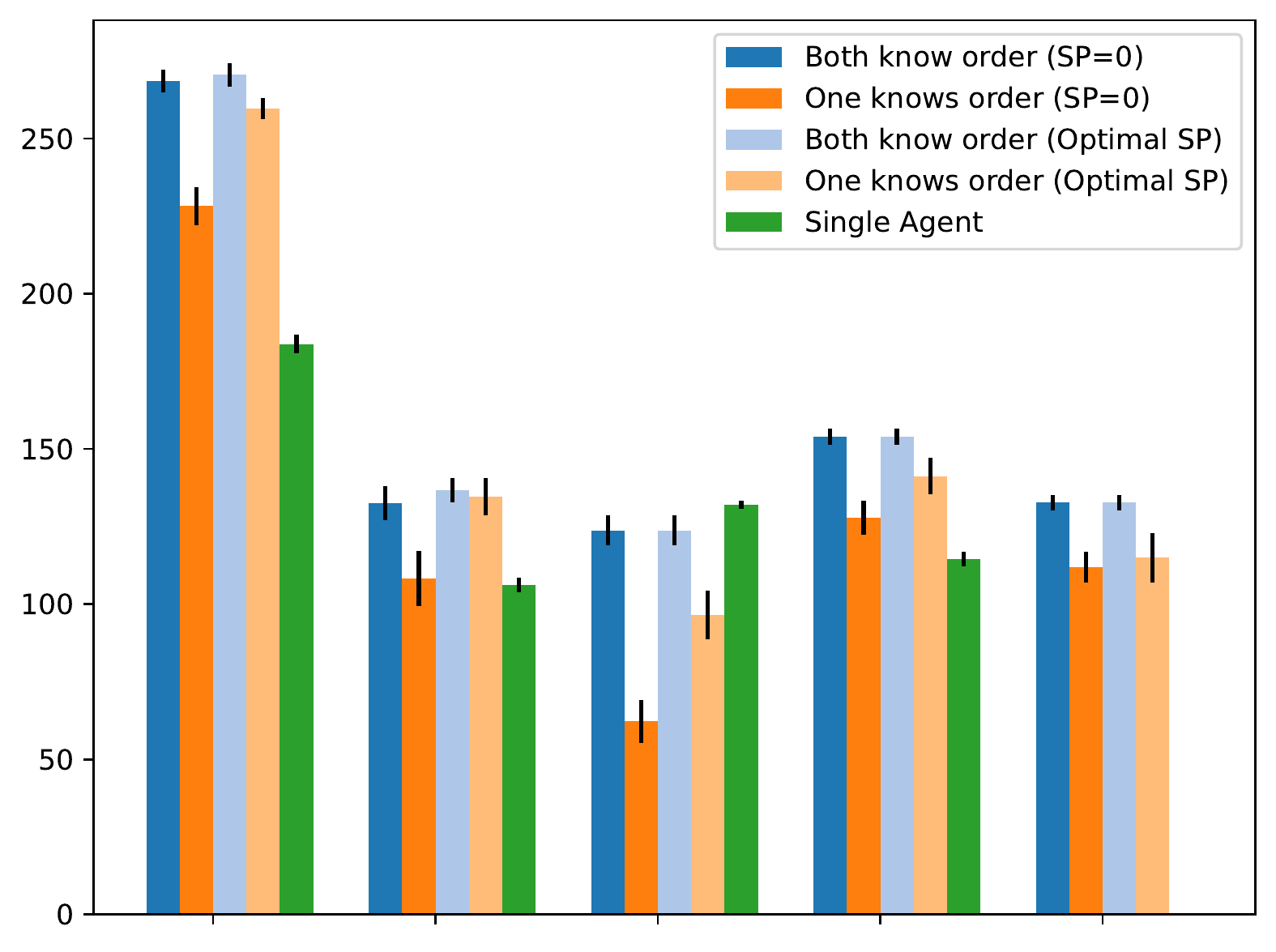}
    \caption{Average reward in the different layouts for a single agent and for two agents with different knowledge (both or only one knowing the orders) and different degrees of belief resonance ($SP=0$ or set to the optimal value). Error bars indicate the standard error.}
    \label{fig:averageRewards}
\end{figure}

Fig.~\ref{fig:averageRewards} shows the average rewards achieved in different kitchen layouts by teams in which a) both agents know the current order vs. only one agent knows it, and b) the agents do not perform belief resonance (SP=0) vs. with the optimal combinations of SP values. To determine these optimal, i.e., best performing, SP values, we tested all combinations between 0 and 1 in 0.1 increments. As a baseline, as well as for the first evaluated aspect, we add the reward achieved by a single agent using the proposed model. This single agent equipped with the proposed model was able to solve the cooking tasks in all but the \emph{forced} layout with fairly consistent performance. The \emph{forced} layout requires two agents, one in each of the separated parts of the kitchen, to be able to prepare any meals at all.

\subsection{Effect of belief resonance when both agents know the orders}

Without any belief resonance (SP=0), the agents achieve scores ranging from 123.75 ($\pm$ 4.75) in the \textit{cramped} layout to 268.5 ($\pm$ 3.55) in the \textit{asymmetric} layout (Fig.~\ref{fig:averageRewards}). Except for the \emph{cramped} layout, two agents performed better than a single one, indicating that the hierarchical model alone already supports some form of uncoordinated collaboration. Considering the optimal combinations of SP values, other than 0 for both, the team performance improves only marginally in the \emph{asymmetric} and \emph{spacey} conditions. 
\emph{asymmetric} improves to 270.5 ($\pm$ 3.89) with SP values of 0 and 0.2 for the two agents and \emph{spacey} improves to 136.75 ($\pm$ 3.93) with SP values of 0.1 for both agents.

\begin{figure}
    \centering
    \includegraphics[width=0.6\linewidth]{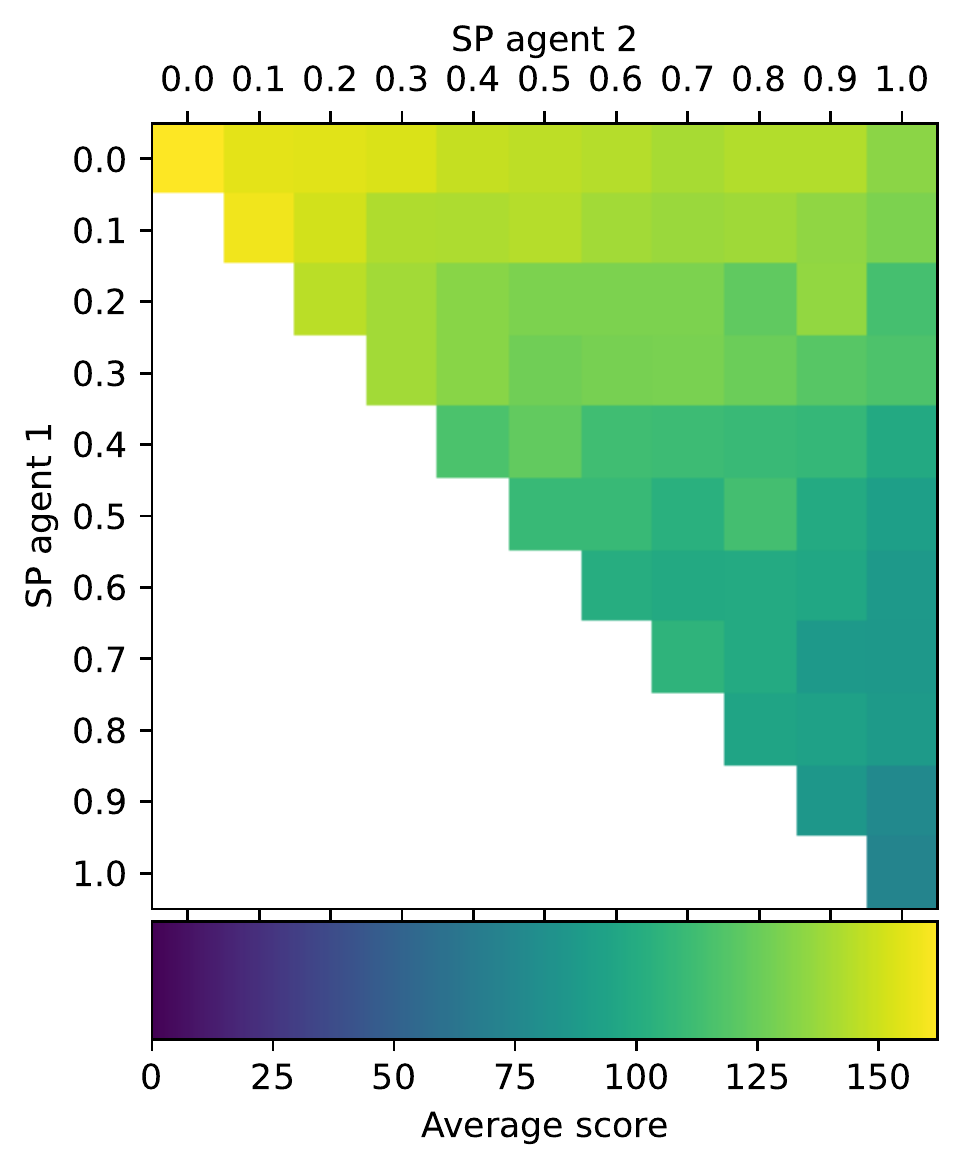}
    \caption{Mean reward for different SP value combinations averaged across the different kitchen layouts when both agents know the orders (averaged across N=20 runs for each layout).}
    \label{fig:SPVariations}
\end{figure}

Fig.~\ref{fig:SPVariations} visualizes the agents' performance depending on their SP value combinations in a heatmap. 
For this, we first computed the average reward for each combination of SP values and layout across the 20 episodes. We then computed the mean score for each combination of SP values. While all combinations were able to complete some orders, the agents performed best with rather low SP values for both agents. 
Considering the layouts individually, we find that 
the optimal combinations for all layouts had one agent not (SP=0) or only minimally (SP=0.1) performing belief resonance. The best SP value was 0.2 for the second agent in \emph{asymmetric}, 0.3 in \emph{cramped}, 0.4 in \emph{ring} and 0.1 in \emph{spacey}. Especially in the \emph{asymmetric} condition, however, other SP values for the second agent achieve similar performance.


\subsection{Effect of belief resonance when only one agent knows the orders}

In this condition, one agent (agent 2) was made ``order blind'' by always receiving a uniform distribution over soup orders.
The average team performance declines when one agent does not know the orders and the agents do not perform belief resonance (SP=0) as can be seen in Fig.~\ref{fig:averageRewards}. However, when considering the optimal SP combinations, the team performance recovers in all but the \emph{forced} layout, often getting close to the performance when both agents are informed of the orders. 

\begin{figure}
    \centering
    \includegraphics[width=0.6\linewidth]{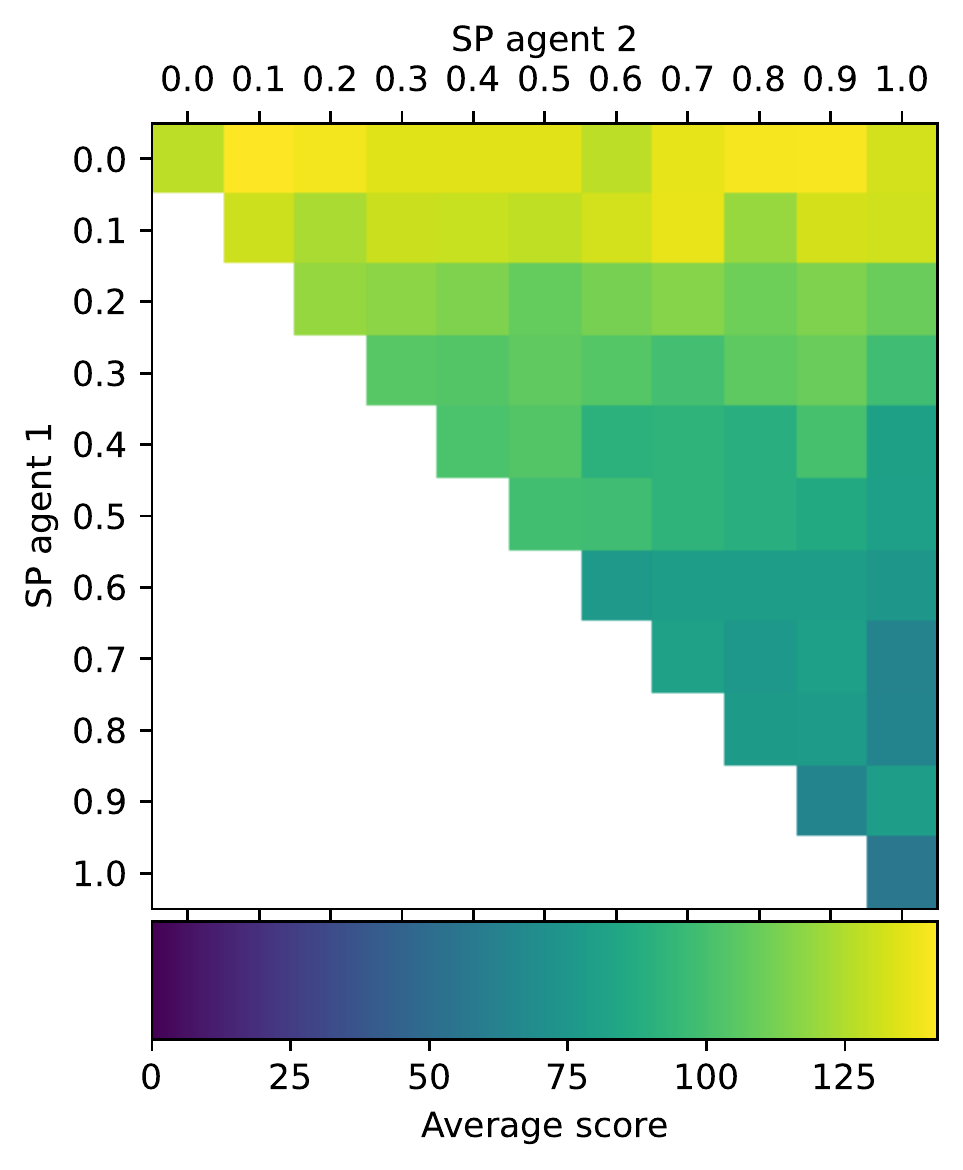}
    \caption{Mean reward for different SP value combinations averaged across the different kitchen layouts when only agent 1 knows the current orders (averaged across N=20 runs for each layout).}
    \label{fig:noOrders}
\end{figure}

As shown in Fig.~\ref{fig:noOrders}, the optimal SP combinations still involve fairly low values for agent 1 that knows the orders (SP of 0 or 0.1). The ``order blind'' agent, however, performs best with SP values between 0.2 (for \emph{spacey}) to 0.9 for (\emph{ring}). Finally, similar to the case when both agents know the orders, performance deteriorates substantially when both agents use SP values beyond 0.6. 

\subsection{Effect of belief integration order}

\begin{figure}
    \centering
    \includegraphics[width=0.6\linewidth]{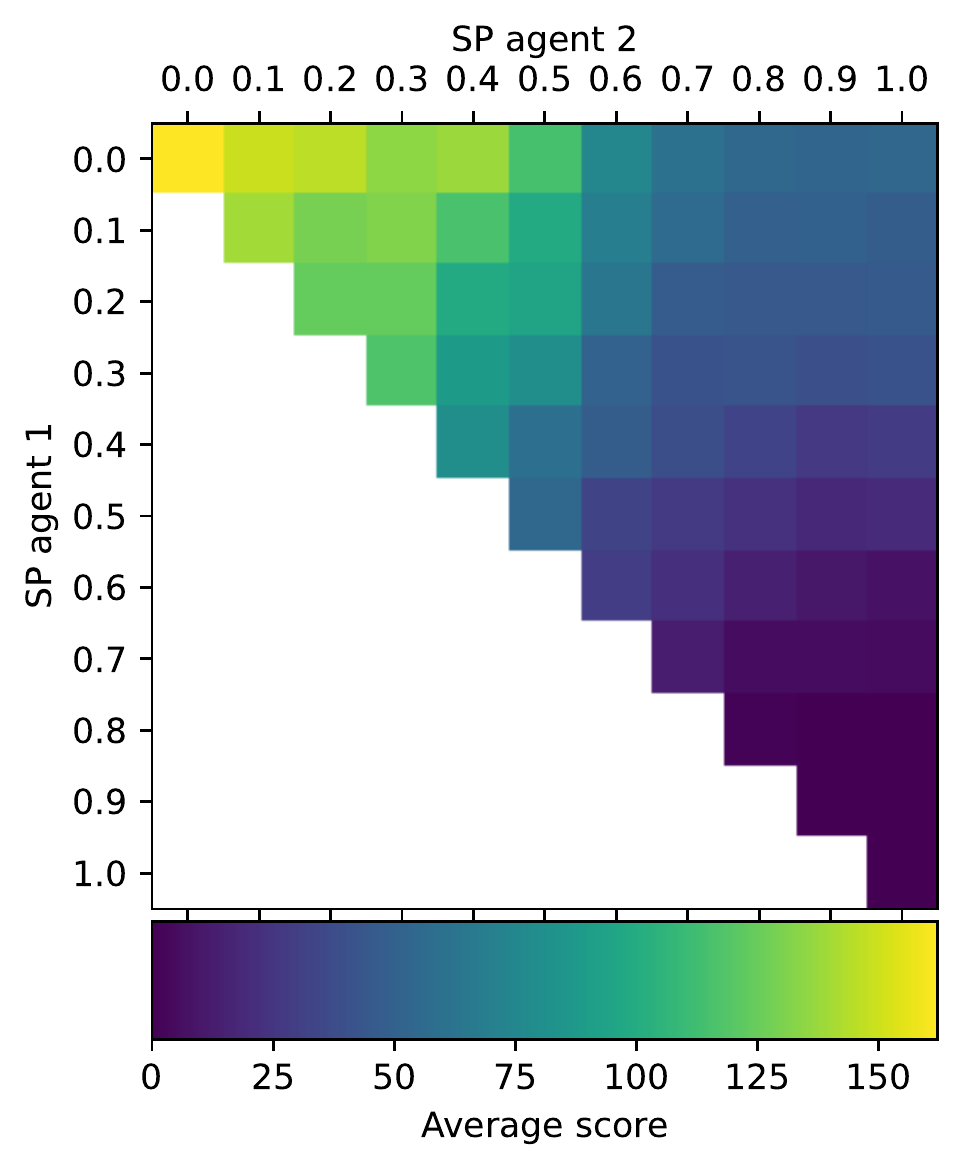}
    \caption{Mean reward for different SP value combinations averaged across the different kitchen layouts when the inferred belief was integrated after the top-down and bottom-up information (averaged across N=20 runs for each layout).}
    \label{fig:ToMIntegrationLast} 
\end{figure}

When we integrate the inferred other-beliefs \textit{after} updating the agent's beliefs using the top-down and bottom-up information (instead of in-between), the average team performance is a lot more dependent on very low SP values, as shown in Fig.~\ref{fig:ToMIntegrationLast}. 
Across all layouts, agents only perform reasonably well with SP values of 0, i.e., without belief resonance. Very low values (0.1 or 0.2) can still yield results comparable to the original integration order presented above. However, team performance quickly breaks down for higher SP values. This is especially the case for the layouts \emph{ring, forced} and \emph{cramped}, where SP values beyond 0.5 for any agent result in the agents mostly failing to complete any orders at all. 

\subsection{Comparison with state of the art: Reinforcement Learning} 

In order to put the results shown in Fig.~\ref{fig:averageRewards} into perspective, we compare them with the results reported by Carroll at al.~\cite{carroll2019utility} for the best RL model in similar kitchen layouts (\emph{asymmetric}, \emph{ring}, \emph{forced}, and \emph{cramped}). Only in the \emph{cramped} layout are the ingredients, the pot and the serving area arranged slightly differently.
It is noteworthy, however, that Carroll et al. used a simpler task by limiting the possible orders to only one type (\textit{Onion} soup). Thus, for comparison, we also restricted the kitchen layouts in Fig.~\ref{fig:kitchen_layouts} to only contain onions. 

\begin{figure}
    \centering
    \includegraphics[width=\linewidth]{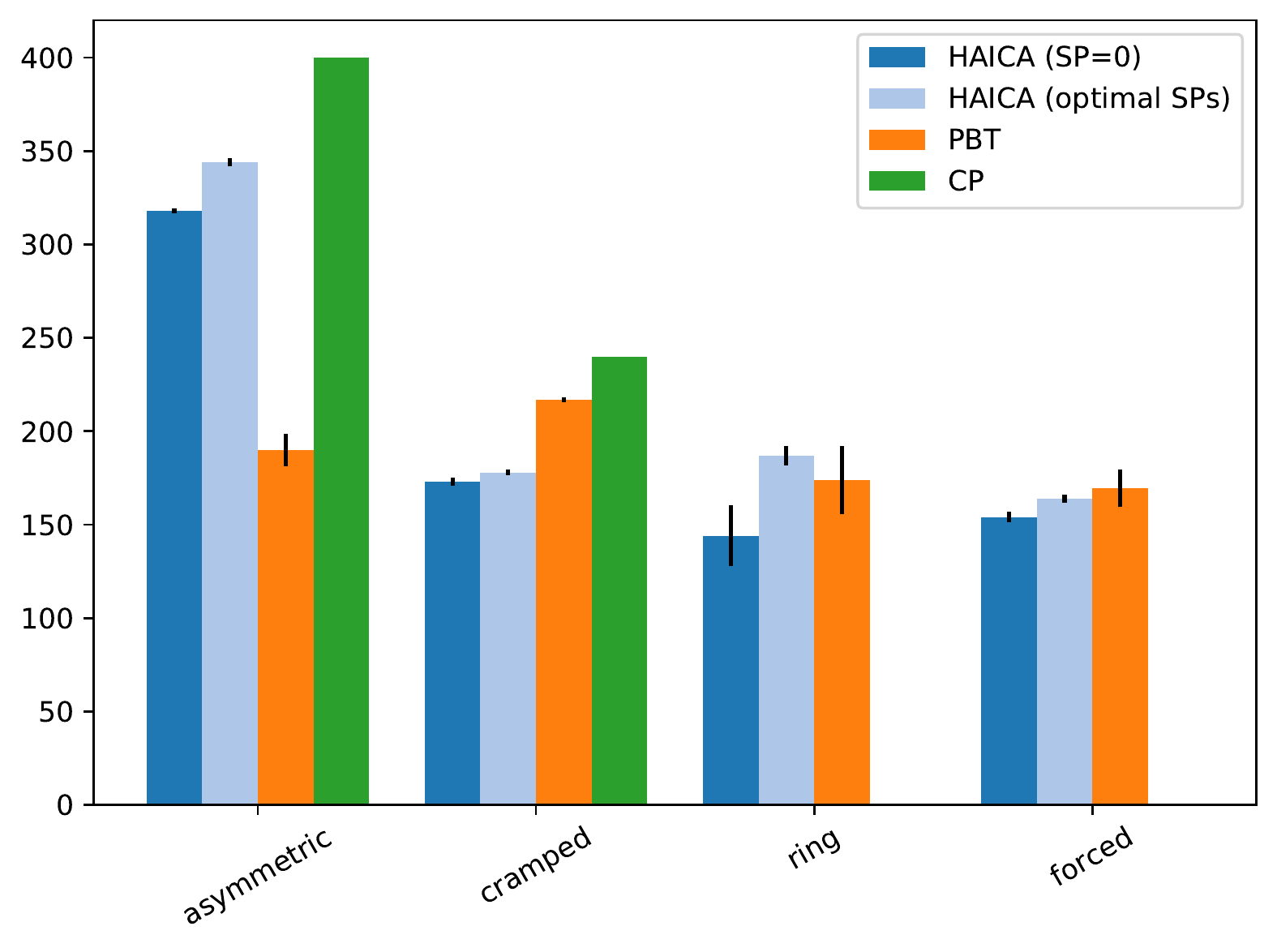}
    \caption{Comparison of average rewards achieved by different agent models in various kitchen layouts (with slight differences in the \emph{cramped} layout).
    PBT represents the deep RL agents trained using population based training, while CP represents the optimal agents using coupled planning presented in \cite{carroll2019utility}. The PBT and CP results are mean scores over 5 trails, while the HAICA scores were averaged over 20. Error bars represent the standard error.}
    \label{fig:carrollComparison}
\end{figure}

Fig.~\ref{fig:carrollComparison} shows the scores obtained with two variants of the HAICA model, alongside the scores reported in \cite{carroll2019utility} for RL agents trained via population based training (PBT). We also added Carroll et al.'s results achieved with an optimal policy determined via coupled planning (CP), in only two of the layouts in which computing the optimal policy was feasible.

The results represent the average reward achieved by the agents after 400 steps, where the \textit{Onion} soup requires 20 time steps to cook and yields 20 points upon delivery. Note that 
Carroll et al.~\cite{carroll2019utility} averaged across 5 random seeds, while the results for the model proposed here were averaged across 20 runs. Note also that we only included the results of agents collaborating with copies of themselves as the performance in \cite{carroll2019utility} generally deteriorated for mixed-agent setups. 

Generally, the performance of the proposed HAICA model is comparable to and only marginally worse than those achieved through deep learning \cite{carroll2019utility} in the layouts \emph{cramped, ring} and \emph{forced}. 
In the \emph{asymmetric} condition, the HAICA models achieve significantly better performance, most likely due to the fact that the agents are able to make use of both pots instead of only one the deep RL agents tend to use. 
Yet, the comparison with the CP agents also demonstrates that HAICA does not reach a score that is achievable with an optimal policy.

\subsection{Comparison with state of the art: Bayesian Delegation}
The mentalizing approach used in the HAICA model with belief resonance is more similar to probabilistic BToM approaches than to RL.
For this reason, we also compared the HAICA model to an explicit planning approach using Bayesian Delegation (BD) \cite{wu2021too}. 
Instead of re-implementing the environment as we did with the soup scenario above, we used the code provided by Wu et al. \cite{wu2021too} and implemented the HAICA agent for their slightly different realization of the Overcooked domain. In this domain, agents are supposed to prepare three different kinds of salads (lettuce, tomato and mixed). 
The preparation of a salad involves cutting the ingredients (lettuces and tomatoes) at a knife station and happens instantly upon interaction.
A major difference is that each object (ingredient or plate) is only available at the quantity  required to complete the order, unlike the dispensers considered above. 
Wu et al. considered three possible tasks: prepare and deliver one \textit{Tomato} salad, two salads, one of each type (\textit{Tomato+Lettuce}) or one \textit{Mixed} salad. 
They further evaluated their model in three different kitchen layouts (Full-Divider, Partial-Divider and Open-Divider) resulting in 9 different scenarios.

Implementing the HAICA agents in this environment is largely identical to the implementation presented above. It basically involves changing the goal domain to the three salad recipes (lettuce, tomato and mixed) and renaming the intentions and their objects according to the new ingredients (lettuce and tomato) and interaction object (cutting board).
While the likelihood functions needed to be adapted to the different environment implementation and dynamics, their functionality remained fundamentally the same as well.

Some precautions are needed in this scenario, though.
Since each ingredient is only available once and they need to be combined, agents can get stuck in both picking up one part of the recipe (e.g., one agent picking up the cut lettuce and the other a plate). 
They may end up cycling through placing an item down for the other agent, and picking it or the other item back up. 
This is not a problem in the soup scenario which included dispensers with unlimited amounts of different items. At the same time, each recipe requires multiple of the same ingredients. 
In order to reduce chances for such cycles, we punished interacting with an item directly after placing it down by reducing the corresponding intention's prior probability.

To compare the HAICA model with the BD model, we use metrics similar to those used by Wu et al.~\cite{wu2021too}. 
Instead of determining a team score after a fixed amount of time, they measured the number of time steps required to complete a single task.
They further evaluated their agents with respect to the number of completed sub-tasks (e.g., cutting the tomatoes for the \textit{Tomato} salad) after a maximum of 100 time steps. 
Since the proposed HAICA model is agnostic to sub-tasks we cannot compare it on that metric. 
Instead, we decided to focus on the percentage of overall successes within 100 time steps. 
An episode was considered a success if the agents managed to complete the task entirely, e.g., only delivering the \textit{Lettuce} salad but not the \textit{Tomato} salad would not count as a success in the \textit{Tomato+Lettuce} task.
Additionally, we measure the actual clock time each episode requires. 
This is important, since our focus is not to be competitive with explicit planning in terms of optimal action selection, but rather to achieve satisficing results that enable real-time interactions.

\begin{table*}
\centering
  \caption{Pairs of SP values that achieved the best success rate in the different scenarios. If two combinations led to the same success rate, the lower number of time steps decided.}
  \label{tab:sp_values} 
  \begin{tabularx}{0.7\textwidth}{cccc}
    \hline\noalign{\smallskip}
    Kitchen Layout & Tomato & Tomato+Lettuce & Mixed\\
    \noalign{\smallskip}\hline\noalign{\smallskip}
    Full-Divider & 0.3 \& 1.0 & 0.8 \& 1.0 & 0.0 \& 0.3 \\
    Open-Divider & 0.0 \& 0.7 & 0.2 \& 1.0 & 0.3 \& 0.4 \\
    Partial-Divider & 0.4 \& 0.4 & 0.3 \& 0.4 & 0.1 \& 0.8 \\
  \noalign{\smallskip}\hline
\end{tabularx}
\end{table*}

\begin{figure}
    \centering
    \includegraphics[width=\linewidth]{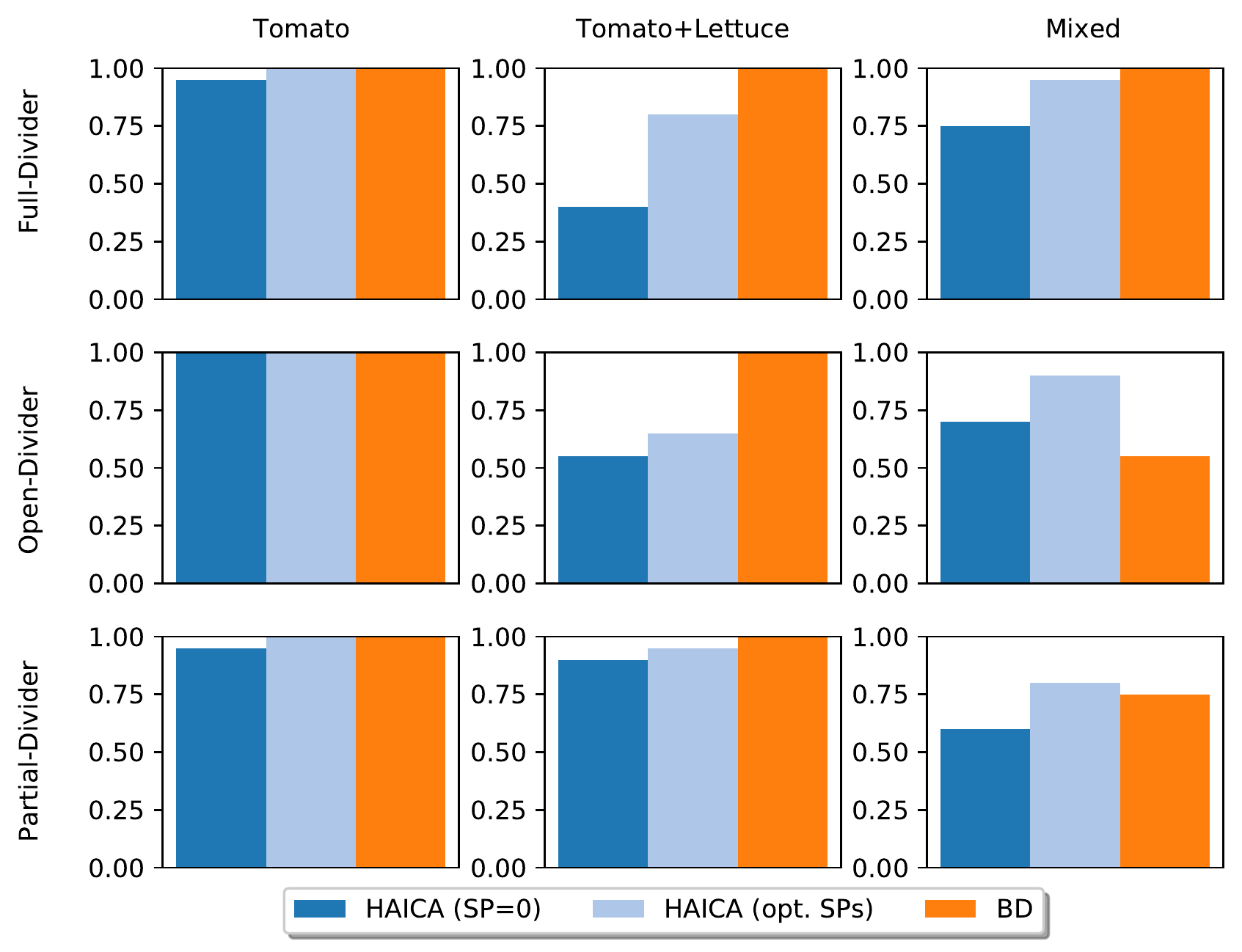}
    \caption{Percentage of successful task completion (out of 20 trials) achieved by different agent model pairs in three different kitchen layouts and three different tasks used by Wu et al. \cite{wu2021too}.
    BD represents the Bayesian Delegation model proposed there, the HAICA models differ with respect to the degree of belief resonance of each agent (no resonance for SP=0, or the optimal degree and SP value for this scenario).}
    \label{fig:wangComparison}
\end{figure}

Fig.~\ref{fig:wangComparison} presents the success percentages for each of the 9 different scenarios achieved by agent teams with two variants of the HAICA model and the BD model. 
The values for the BD model were obtained by running the replication script provided by Wu et al.~\cite{wu2021too} without any changes except for an additional time-clock measurement. We also re-used their script to evaluate the HAICA model by simply replacing their BD agent model with the respective HAICA model. 
That way, we evaluated all combinations of SP parameters between 0 and 1 in 0.1 increments for the HAICA agents.
We present here the results for pairs with SP=0 as well as the optimal SP values, i.e., those that maximize the agent's success rate, which are listed in Table \ref{tab:sp_values}. 
The results show that all agent teams always succeed in preparing just the \textit{Tomato} salad, except those using the HAICA model without belief resonance (SP=0). When preparing two dishes (the \textit{Tomato+Lettuce} task), the HAICA models always perform worse than BD. On the other hand, BD appears to struggle most with the \textit{Mixed} salad in the Partial-Divider and Open-Divider kitchen layouts. In this task, the HAICA model with optimal SP parameters performs better than BD.
Comparing these optimal SP values to the soup domain discussed before, we see they are notably larger in this domain. This is especially true for the \textit{Tomato+Lettuce} task in two of the three kitchen layouts.

Table~\ref{tab:wangComparison} lists further results using comparable metrics. Values are aggregated across all 9 scenarios 
used in \cite{wu2021too}, where each scenario was re-run 20 times with different random seeds. 
The percentage of successes are averaged over these 20 seeds as well as the different scenarios. The other statistics are averaged across all episodes.
BD relies on Q-value estimations computed using BRTDP \cite{mcmahan2005bounded}. To use BRTDP, one configures the maximum number of steps to simulate along with the maximum number of simulations to perform when estimating the Q-values. In order to compare BD with different computational budgets, we also tried running BD with less optimal but computationally cheaper Q-value estimations. To that end, we decreased these parameter values from 75/100 (default) to 10/2 and even 1/1 (for \#steps and \#simulations, respectively). 
All models were run on the same machine using an Intel(R) Xeon(R) CPU E5-1620@3.50GHz.

\begin{table*}
\centering
  \caption{Average number of steps, percentage of successes, average time per episode (in seconds) and average time per step (in seconds) with their standard error for two variants of HAICA with different SP values and the Bayesian Delegation (BD) model proposed in \cite{wu2021too} using different parameters for the Q-value approximation with BRTDP (maximum number of steps per simulation/maximum number of simulations). The default values are 75/100.} HAICA with SP=0 does not use any belief resonance, while optimal (opt.) aggregates the best SP combinations for each layout. (Averaged across 20 seeds and the 9 scenarios.)
  \label{tab:wangComparison} 
  \begin{tabularx}{\textwidth}{cccccc}
    \hline\noalign{\smallskip}
    Model & Time Steps & Success & Run-time & Time per Step\\
    \noalign{\smallskip}\hline\noalign{\smallskip}
    HAICA (SP=0) &  71.41 $\pm$ 1.78 & 0.76 $\pm$ 0.03 & 1.54 $\pm$ 0.06 & \textbf{0.02 $\pm$ 0.00} \\
    HAICA (opt. SPs) & 60.79 $\pm$ 1.69 & 0.89 $\pm$ 0.02 & \textbf{1.32 $\pm$ 0.05} & \textbf{0.02 $\pm$ 0.00} \\
    BD (default) & \textbf{36.52 $\pm$ 1.51} & \textbf{0.92 $\pm$ 0.02} & 2111.55 $\pm$ 133.85 & 55.87 $\pm$ 3.33  \\
    BD (10/2) & 51.02 $\pm$ 1.85 & 0.86 $\pm$ 0.03 & 67.14 $\pm$ 2.95 & 1.29 $\pm$ 0.02  \\
    BD (1/1) & 77.12 $\pm$ 1.93 & 0.53 $\pm$ 0.04 & 21.40 $\pm$ 0.77 & 0.27 $\pm$ 0.01  \\
  \noalign{\smallskip}\hline
\end{tabularx}
\end{table*}

These results show that the approach presented here is significantly worse compared to the optimal planning achieved by BD with default parameters when it comes to the number of time steps (or actions) that are required to solve the task.
As an episode is deemed a failure after 100 time steps, the percentage of successes for this domain is also worse, especially in the case with $SP=0$ for both agents, i.e., without belief resonance.
However, if we aggregate the data for the best SP combinations for each of the different layouts, the HAICA model performs almost as well as BD in terms of successes.

The differences in performance can be put into relation with the differences in run-time. While the agent pair equipped with HAICA models completes an episode in 1.37-1.55 seconds on average, agents using the BD model with optimal planning require more than 35 minutes to complete an episode on average. 
Naturally, this translates directly to the time the agents require to decide on each action. 
While each step, i.e., the action selection of both agents and the update of the environment, took only 0.02 seconds with the HAICA models, it took close to 56 seconds on average with the BD models.
If one neglects the time to update the environment (which is identical regardless of the chosen agent model), each BD agent takes around 28 seconds to decide upon an action on average, while each HAICA agent only requires around 0.01 seconds.
BD's real-time capability can be greatly improved by limiting the computational resources of the Q-value approximation. A reduction from 75/100 of simulated steps and repeated simulations to 10/2 provides an impressive speedup, resulting in around 1.29 seconds per action or even 0.27 seconds per action by only simulating 1 step only once. However, the worse Q-value approximation results in a lower number of successes and a related increase in the number of steps required to complete each scenario. Notably, even with minimal Q-value estimation, BD requires substantially longer than HAICA. 

\section{Discussion}
Our evaluation studies reveal several interesting insights about the proposed HAICA model.
First, the results show that the agents are able to collaboratively solve the tasks of preparing meals in the different environments in the ``soup'' domain, albeit with varying success. 
The HAICA models performed best in the \emph{assymetric} layout. 
However, the raw scores cannot be compared directly across kitchen layouts.
One important factor for the differences between kitchen layouts is the distance between ingredients and the pots.
The number of steps that agents had to perform to move an ingredient into a pot is the shortest in the \emph{asymmetric} layout.
This layout also has the benefit that agents cannot block each other's path in this condition.

Notably, and contrary to our expectation, the agents were also able to solve layouts that necessitate a certain amount of collaboration such as \emph{ring} or \emph{forced}. They did so even without belief resonance (SP=0). 
This can be explained by the chosen intentions and their likelihood functions. 
The \emph{hand-over} intention will become active when the agent is not able to do anything else that may help it complete the meal, regardless of what the other agent may currently be doing. 
That way, the agent on the l.h.s. in the \emph{forced} layout (see rightmost sub-figure in Fig.~\ref{fig:kitchen_layouts}) will pass over potentially required items. 
Likewise, the agent on the r.h.s. may then be forced to work on an order it did not initially intend to, simply because it only has access to the ingredients for the other order. 
It is worth noting that this does not imply explicit coordination. The agents simply realize from their bottom-up affordances that all they can do to achieve their goal is to use the \emph{hand-over} action.
When considering the optimal combinations of SP values, no significant improvements were achieved, with only minor improvements in the \emph{asymmetric} and \emph{spacey} layouts. The best SP combinations usually include very low SP values.
This is most likely due to ``collaboration'' being inherent in the intentions such that
the agents start collaborating simply by doing what they are able to do given the nature of the task and their joint situatedness in the shared environment.

Belief resonance does have a very positive effect in scenarios with unbalanced information. 
When only one agent knows the current order, belief resonance is sufficient for the other agent to take up a follower role in the collaboration.
Consequently, team performance suffered substantially without belief resonance (SP=0), while the agents' performance almost recovered to the level of the first condition when the agents used optimal SP values for most of the kitchen layouts. 
The optimal combinations usually contained one ``leader'' with a low SP ($\leq 0.1$) and one  ``follower'' with values going up to 0.9, i.e. becoming very susceptible to the inferred intentions and goals of the other. 
Only in the \emph{forced} layout there is no improvement.
We suspect that this stems from the fact that the agent on the right-hand side usually does not get much of a choice about its next actions regardless of it knowing the orders or not.
The agent on the left-hand side would just take random guesses if it does not perceive the orders and is unable to infer useful information from the other agent. 
Overall, these results indicate that belief resonance can be a useful building block for fast and adaptive collaboration through on-the-fly coordination, especially in situations that benefit from leader-follower dynamics and that provide agents with shared action affordances.

When looking at the SP values in more detail, we find that in the soup domain large SP values for both agents usually lead to a reduced team performance, in the worst case with agents being unable to complete a single order. 
This is understandable because agents with very high SP values overwrite their own goals and intentions at every step with those inferred about the other agent. 
While this may seem like an argument against belief resonance, we would interpret it as a need for adaptive collaboration and complementary roles \textit{in addition} to it. 
A fixed SP value can be detrimental as it may lead to situations where the agents are too susceptible to the inferred beliefs, resulting in both agents trying to ``follow the lead'' of the other agent with no one leading. 
Indeed, the results show that one competent agent leading (and being more ``stubborn'') is usually better. 
Further, apart from using lower SP values, one could also avoid this problem by providing agents with memory and having them stick to chosen intentions unless those are completed or become impossible.
Future work should look at dynamically adapting the SP parameter in order to naturally take up leader or follower roles depending on the other agent. 

Regarding the order in which information is integrated during belief resonance, we find that performance is a lot worse for all but only small SP values when the inferred other-beliefs are integrated into self-beliefs last. 
This can be explained when considering the meaning behind the different sources of information.
Bottom-up information reflects the agent's current affordances. 
By suppressing or overwriting its own beliefs with the inferred beliefs (depending on the SP value), an agent may end up choosing active intentions that are not possible for itself. 
When using the order proposed in Eq.~\ref{eq:tom_belief_update_0} and \ref{eq:tom_belief_update}, however, the inferred belief is first combined with the top-down information, before being ``validated'' against bottom-up information about current action affordances of the environment.

Although HAICA is not designed to search for the best possible collaborative actions, we were surprised by its performance compared to other state of the art approaches. The model achieves a comparable performance as state of the art deep RL approaches \cite{carroll2019utility} trained and tested on simpler versions of four of the layouts considered here. 
In the \emph{asymmetric} layout, HAICA agents even outperformed the RL agents, which tend to only use one of the available pots and wait for the soup to be fully cooked, instead of preparing ingredients in the second pot in the meanwhile. 
The proposed HAICA model instead tries to work on present orders as long as the bottom-up affordances allow appropriate actions. These agents thus make use of the second pot while a soup is cooking in the first.
The three other kitchen layouts either had only a single pot to begin with (\emph{cramped}), or the position of the second pot was not accessible, at least when the first one was in use (\emph{ring} and \emph{forced}).
It is also worth noting that, due to the simplification of having only one order  \cite{carroll2019utility}, HAICA could only play out its mentalizing and belief resonance at the intention layer. 

Considering the comparison with Bayesian Delegation (BD) \cite{wu2021too} in the different salad setting, we find that the HAICA model selects less optimal actions. This is not surprising as BD agents perform explicit planning and try to find the optimal plans given their sub-task assignments. This is similar to the optimal behavior achieved by coupled planning in \cite{carroll2019utility}.
The minimalistic action selection strategy employed here does not take the other agent's future actions into account when planning own actions. 
The environment's affordances may cause each agent to pick up one of the items that need to be combined.
While the agents realize that they need to hand their item over to the other agent, since they cannot complete the recipe by themselves, they do not consider the other agent potentially doing the same thing. 
Since there is no explicit coordination, both agents tend to get ``stuck'' placing their item down and picking it back up or taking the item the other one just put down. 
This leads to many unnecessary actions and agents only have a chance to get off this cycle due to the proposed ``punishment'' of these repetitive intentions.
In order to detect and actively avoid such scenarios, more explicit planning or fine-tuning would be required. 

Interestingly, the salad scenario appears to allow for SP values that are substantially higher than in the soup domain. 
While the optimal roles still included one ``leader'' agent with a relatively low SP value in most scenarios, both ``leader'' and ``follower'' agents had higher SP values overall. This is probably due to the fact that the agents only need to complete one dish. Therefore, there is no ambiguity with respect to the agent's goal in this domain.  
This would already reduce the effect belief resonance can have on the agent's behavior.
Another reason could be that each object is available only as often as it is required.
This reduces the effect of belief resonance even further as the inferred intention of the other will often directly be invalidated by the bottom-up information, since it is no longer possible for the other agent to perform the same intention.
Yet, while the SP value did not appear to have much of an effect on the agent's success rate in both the \textit{Tomato} and \textit{Tomato+Lettuce} tasks, agents with lower SP values performed better when working on the \textit{Mixed} salad tasks.
These findings provide additional evidence for the need for artificial agents to be able to socially adapt their role and collaborative behavior depending on the situation as well as the other agent. In the HAICA model, this adaptation can be operationalized directly by means of adjusting the degree of belief resonance, i.e., the SP parameter dynamically.

Finally, we have argued that real-time collaboration hinges on the ability of agents to coordinate with others efficiently and on-the-fly. In this regard, HAICA bears a significant advantage over other approaches such as the recent BD model. While BD agents can select actions to prepare the dishes in only a minimal number of steps, the high computational costs of decision making on average lead to almost half a minute of time needed to decide on which action to perform. As a result, the agents required much longer to deliver the salad than the proposed HAICA model, which selects its action in a fraction of a second. BD's efficiency can be greatly improved at the cost of the accuracy of the Q-value estimation and, in turn,  team performance. 
However, the time differences between BD and HAICA are not solely a product of BD's Q-value estimation. Even with only minimal estimation, BD still requires more than 10 times longer than HAICA while performing significantly worse at that point. 
This difference likely stems from BD's explicit planning approach that tries to find the best possible actions for all agents. 

While efficiency is not a concern in the synthetic scenarios discussed above, other real-world scenarios or domains do impose time constraints on the agents. For example, a real-world kitchen domain would have dynamics that do not wait for the agents' actions, such as a meal burning if cooked too long. 
Another aspect where real-time capability is crucial is for human-agent interaction. 
Not only would a human collaborator get frustrated with an agent that reacts too slowly, but the human would likely perform multiple actions in the time the agent still considered how to best respond to the first action it observed. This may result in the agent planning with an invalid world state and may thus lead to serious collaboration problems.

\section{Conclusion}

In this paper we have started to explore if ``satisficing collaboration'' can emerge between agents that produce highly situated, prediction-based behavior while being affected by what they assume other agents intend to do. We have proposed a model for \textit{hierarchical active inference for collaborative agents} (HAICA), which integrates ToM-inferred mental states of another agent into a probabilistic and predictive-processing based goal and intention formation. Akin to how sensorimotor processes start to resonate with observed behavior in the human or animal brain, we call this mechanism \textit{belief resonance}. Crucially, HAICA does not require any specific means of separating the mental perspectives of the two agents, or for explicitly planning and coordinating their activities. It thus provides a rather minimal ``cognitive infrastructure'' for collaborative agents and is to be seen as one building block that needs to be complemented with more powerful cognitive processes for inference and planning. Our goal was to explore what kinds of collaborative tasks can be tackled with such a minimal account and how efficiently and robustly this would work.

Simulations with fully implemented versions of HAICA in different versions of the Overcooked domain show that two agents equipped with the proposed model were able to prepare meals together in a resource-efficient manner. We saw coordination between agents \textit{emerge} while they dynamically adapted to each other, without engaging in explicit coordination or global planning. That is, the  coordination was mediated only through (1) their co-situated perceiving and acting in a shared world, and (2) their ability to understand and resonate with each other's action in terms of the intentions and goals that drive it. 
The impact of belief resonance, from the point of view of the belief-resonating agent, is thereby controlled by a susceptibility parameter (SP) that implies different emerging roles of the agent (leader or follower). This leads to a coordination effect that is most notable in situations where the agents have different access to task-related information (here, knowing the current orders).

We want to reiterate that we do not see the proposed model as a general solution to the collaboration problem. Although our results are comparable to the state of the art in deep reinforcement learning models \cite{carroll2019utility}, they are, of course, worse than those achieved when considering optimal (joint) policies \cite{wu2021too}. 
However, determining such a policy is infeasible for independent agents and is computationally very demanding even in simple 2D domains as considered here. 
Enabling collaboration without dedicated prior coordination thus requires agents to adapt to each other quickly, continuously and effectively. 
Such an adaptation may involve explicitly planned coordination or no coordination at all depending on the situation. 
Real-world collaborative systems (e.g., collaborative robots) will likely need to adapt their own coordination 
approaches depending on the situation and the other agent.
That is, agents may need to switch between deliberating about (planning) the most suitable actions and reacting more ``intuitively'' to time critical changes. The results presented here provide valuable insights into possible building blocks for this based on which situated systems with more sophisticated collaboration skills can be built. 

Different improvements are possible to reduce or prevent current shortcomings of the presented model, such as not taking the other agent's likely next actions into account when deciding on low-level actions, or not holding on to selected intentions. 
Such improvements may 
involve more explicit reasoning and planning. 
One could, e.g., use the explicit BRTDP planner used in \cite{wu2021too} instead of our best-first search realization.
If the resulting future expected reward estimates are also used to compute the action likelihood for the ToM module, one would create a HAICA model with explicit planning very similar to BD.
We would expect such a model to be more competitive with respect to the optimality of the produced behavior.
One should, however, always take the computational burdens of more sophisticated methods into account, so as not to impede the overall system's real-time capabilities. 
Especially if we consider real-world applications and/or human-agent interactions, the real-time capabilities of the developed collaborative agents becomes crucial.
Furthermore, we have only tested the HAICA model in scenarios with two agents. Future work should look into how to apply HAICA to interactions involving multiple agents. One would likely need to implement attention mechanisms that select the agent to resonate with.
Another important aspect for future work is the need to dynamically determine and adapt an agent's role in complex collaboration scenarios. 
Dynamic role adaptation may in part be achievable with a minimal model, e.g.,
by allowing belief resonance to be variably susceptible. 
In this sense, the model presented here is a first step towards elucidating how collaborative behavior can emerge from the dynamic interplay of prediction-based perception and action, minimal social cognition, and situational affordances.

\bibliographystyle{spmpsci}
\bibliography{bibliography}

\end{document}